\documentclass[aip,pop,amsmath,amssymb,reprint]{revtex4-1}

\usepackage[dvipsnames]{xcolor}
\usepackage{graphicx}
\usepackage{dcolumn}
\usepackage{bm}
\usepackage{epstopdf}
\usepackage{natbib}
\usepackage{subcaption}
\usepackage{tabularx}
\usepackage{mathrsfs}
\usepackage{upgreek}

\usepackage[colorlinks=true, allcolors=blue]{hyperref}
\usepackage[capitalise]{cleveref}
\usepackage[a]{esvect}
\setcitestyle{square}
\usepackage[toc,page]{appendix}
\usepackage{array}
\newcolumntype{P}[1]{>{\centering\arraybackslash}p{#1}}

\usepackage[font={small,normalfont}]{caption}
\newcommand{\micron}{{\upmu\mathrm{m}}}

\definecolor{darkred}{rgb}{0.55,0,0}
\definecolor{editgreen}{rgb}{1,1,1}

\newcommand{\AVA}[1]{\textcolor{black}{#1}}

\begin{document}
\title{Longitudinal-field particle acceleration in channels generated by strong electromagnetic waves in pair plasma}

\author{Kavin Tangtartharakul}\thanks{Corresponding author: ktangtar@ucsd.edu}
\affiliation{Department of Mechanical and Aerospace Engineering, University of California San Diego, La Jolla, CA 92093, USA}

\author{Andrei M. Beloborodov}
\affiliation{Physics Department and Columbia Astrophysics Laboratory, Columbia University, NY 10027,USA}

\author{Carolina Herrera Hernandez}
\affiliation{Department of Mechanical and Aerospace Engineering, University of California San Diego, La Jolla, CA 92093, USA}

\author{Mihail O. Cernaianu}
\affiliation{ELI-NP, “Horia Hulubei” National Institute for Physics and Nuclear Engineering, 30 Reactorului Street, RO-077125 Bucharest-Magurele, Romania}

\author{Alexey Arefiev}
\affiliation{Department of Mechanical and Aerospace Engineering, University of California San Diego, La Jolla, CA 92093, USA}

\date{\today}

\begin{abstract}
Relativistically strong electromagnetic waves interacting with pair plasmas appear in astrophysical settings and may become accessible in future laboratory experiments using high-power lasers. Transverse modulations and filamentation can turn extended wavefronts into structures of a finite transverse width. We examine the interaction of such a finite-width wave incident on a cold, uniform pair plasma using a three-dimensional kinetic particle-in-cell simulation. The wave does not penetrate the plasma and instead advances by creating a fully evacuated channel. The wave loses energy at a rate that cannot be explained by plasma clearing alone. We find that the excess losses are caused by strong particle acceleration near the walls of the channel. The acceleration is driven by a longitudinal electric field associated with the transverse wave structure and the geometry of the self-generated channel.
\end{abstract}

\maketitle

\section{Introduction}

In astrophysics, electron-positron plasmas and strong electromagnetic waves are found around compact objects. In particular, neutron stars produce intense radio waves, including extremely powerful fast radio bursts (FRBs)~\cite{Petroff.AAPR.2022.FRB}. In the laboratory, there is strong interest in using high-power lasers to produce relativistic electron-positron pairs and to develop platforms for studying pair-plasma processes relevant to high-energy astrophysics~\cite{Chen.PoP.2023.Perspectives}.

Pair plasmas respond to electromagnetic waves in a qualitatively different way from electron-ion plasmas. The key difference is the mass symmetry of the two species. Electrons and positrons have the same mass, so the wave-driven motion of the two species is symmetric. As a result, the strong charge-separation electric fields familiar from electron-ion plasmas are suppressed. The wave can still drive strong currents and transfer energy and momentum to the plasma, but the absence of an ion-electron mass disparity changes the plasma response.

Plane-wave models demonstrate basic features of this interaction. In particular, recent work has shown that a relativistically strong plane wave incident on a cold pair plasma can be completely reflected even when the plasma is underdense according to the usual linear cutoff criterion~\cite{Tangtartharakul.PRL.2026.Reflection, Sridhar.PRL.2026.Interaction}. In this regime, the reflected wave deposits momentum into the plasma and drives a moving reflection front. The plane-wave picture is useful because it provides analytic estimates for the front velocity, the compressed plasma density, and the energy transferred to the swept-up particles.

A more realistic picture of wave-plasma interaction may include transverse modulations and filamentation instability of the wave~\cite{Max.PRL.1974.SelfFocusing,Langdon.PRL.1975.Filamentation}. These effects can create wave structures of a finite transverse width and allow plasma to leave the interaction region transversely. Once the wave has a finite transverse width, the basic questions have to be revisited: how does the wave transfer energy and momentum to the plasma, how is the transferred energy distributed among the plasma particles, and what controls the depletion length of the wave?

In this paper, we isolate the regime of a wave with finite transverse width using a three-dimensional kinetic particle-in-cell simulation of a cold pair plasma irradiated by a relativistically strong, \AVA{linearly polarized} electromagnetic wave. For the plasma density and wave strength considered here, the wave does not propagate through the pair plasma and instead advances by creating a fully evacuated channel (while an electron-ion plasma at the same electron density would be relativistically transparent). We find that the wave loses substantially more energy than required to clear the channel and show that this excess depletion is caused by particle acceleration in the longitudinal electric field inside the self-generated channel.

The paper is organized as follows. \Cref{sec:channel_formation} describes channel formation and compares it with the plane-wave picture. \Cref{sec:estimates} estimates the energy cost of channel clearing. \Cref{sec:energetics} shows that channel clearing cannot account for the observed wave depletion. \Cref{sec:channel_acc} identifies the longitudinal electric field as the dominant source of energy gain for the energetic particles. \Cref{sec:energy_gain_estimate} provides estimates for this energy gain and connects them to the channel geometry. \Cref{sec:summary} summarizes the main findings and discusses their broader implications.

\clearpage \newpage

\begin{figure*}[t]
    \centering
    \includegraphics[width=0.95\linewidth]{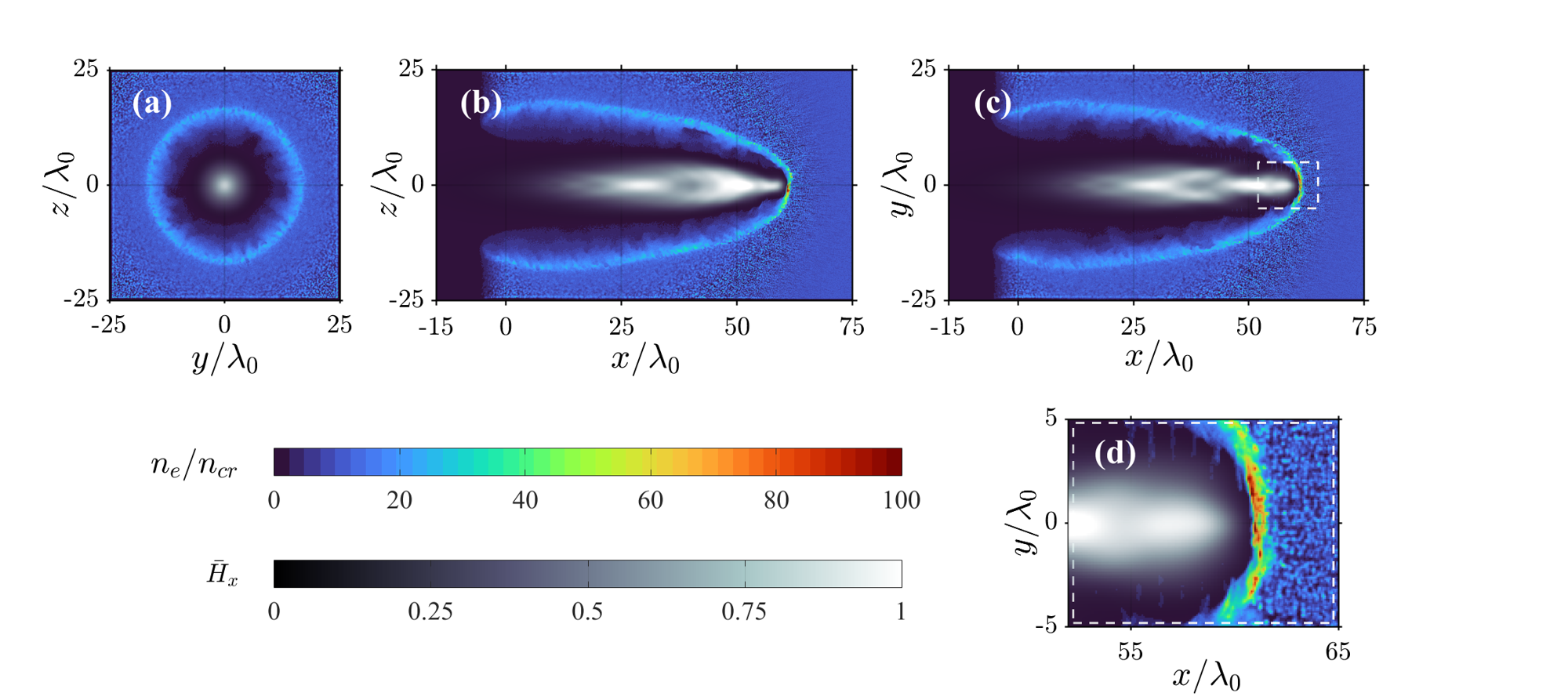}
    \caption{Electron density and wave-field envelope at $t/\tau = 130$. The color scale shows $n_e/n_{cr}$ in three orthogonal slices: (a) $x/\lambda_0 = 25$, (b) $y/\lambda_0 = 0$, and (c) $z/\lambda_0 = 0$. Panel (d) magnifies the dashed region in (c). The grayscale overlay shows the normalized smoothed envelope obtained from the transverse field $E_y$. \Cref{app:field-overlay} provides details on how the envelope is computed, smoothed, and normalized. The envelope also sets the overlay opacity, where 0 is fully transparent and 1 is fully opaque.}
    \label{fig:density}
\end{figure*}

\section{Self-consistent channel formation}
\label{sec:channel_formation}

We refer to a wave as relativistically strong when its normalized amplitude $a_0$ exceeds unity:
\begin{equation}
a_0 = |e|E_0/mc\omega_0 > 1,
\end{equation}
where $E_0$ is the electric field amplitude, $e$ is the electron charge, $c$ is the speed of light, and $\omega_0$ is the wave frequency. At $a_0>1$, the motion of an individual electron or positron in the wave becomes relativistic. The wave magnetic field can then induce a strong longitudinal force and this is why a relativistically strong wave can effectively push pair-plasma particles forward.

We first review the case of a plane wave. The field geometry of a plane wave is such that it can only push a plasma forward. This constraint leads to a striking result: a relativistically strong electromagnetic wave can fail to propagate even when the initial density $n_0$ of each species is below $n_{cr}/2$, where $n_{cr}=m\omega_0^2/4\pi e^2$ is the usual critical density, or cutoff density, for a linear electromagnetic wave in a cold electron-ion plasma. For a cold uniform pair plasma irradiated by a circularly polarized wave with normalized amplitude $a_0 > 1$, the transparency threshold~\cite{Tangtartharakul.PRL.2026.Reflection} is
\begin{equation}
n_0/n_{cr} \approx 1/(32a_0^2).
\label{eq:1D_threshold}
\end{equation}
Above this threshold, the wave does not propagate into the plasma. It is helpful to introduce the plasma frequency defined as $\omega_p = (4\pi n_0e^2/m)^{1/2}$, so that the threshold condition becomes $\omega_p/\omega_0 \approx 1/(\sqrt{32}a_0)$.

The threshold behavior is not a property of the unperturbed plasma, but rather a consequence of the self-consistent evolution driven by the wave. As the wave compresses the pair plasma, weak reflection seeds density spikes that form a moving Bragg-like grating. The enhanced reflection then enables a transition to a regime where the plasma--vacuum interface sustains complete reflection. The reflective boundary (hereafter called the reflection front) moves into the plasma with average velocity $v_b$ given by (see \cref{sec:app:rel-hole-boring-lp})
\begin{equation}
v_b/c \equiv  \beta_b = \frac{1}{1+\sqrt{\psi}},
\label{eq:beta_b_channel}
\end{equation}
where $\psi \equiv 2n_0/(a_0^2 n_{cr})$ for a circularly polarized wave, and
\begin{equation}
\psi = 4n_0/(a_0^2 n_{cr})
\end{equation}
for a linearly polarized wave -- the case studied in the present paper. At $\psi \ll 1$, velocity $v_b$ is relativistic and the corresponding Lorentz factor is large:
\begin{equation} \label{gamma_b}
\gamma_b = (1-\beta_b^2)^{-1/2} \approx \frac{1}{ \sqrt{2}} \psi^{-1/4} \gg 1.
\end{equation}
In this regime, the density at the reflection front greatly exceeds the original unperturbed density $n_0$. The reflection front then produces a compressed plug that moves forward with the front. The density of each species in the plug is
\begin{equation}
n_{\rm{plug}} = 2 \gamma_b^2 n_0 \approx n_0 \psi^{-1/2}.
\label{eq:plug_density}
\end{equation}

We now relax the plane-wave approximation by considering a wave of finite transverse width. The plane-wave result provides a useful baseline, but a finite-width wave can clear plasma from its path in a qualitatively different way. Unlike a plane wave, a finite-width wave allows the plasma to leave the interaction region in the transverse direction as well. This is a key distinction because transverse expulsion changes both the plasma structure and the propagation of the wave.

To illustrate the structure produced by a finite-width wave, we performed a kinetic 3D particle-in-cell (PIC) simulation. In this simulation, a relativistically strong, linearly polarized electromagnetic wave is normally incident onto a cold pair plasma from vacuum. The incident wave is a Gaussian beam. In the absence of the plasma, it would focus at $x = 0$, where the normalized amplitude would reach its peak value of $a_0=50$ at $t = 92.5 \tau$. Here, $\tau = 2 \pi / \omega_0$ is the wave period and $t = 0$ is the start of the simulation. The beam is relatively narrow, with a waist size of $5\lambda_0$, where $\lambda_0 = 2 \pi c / \omega_0$ is the vacuum wavelength. This deliberate choice emphasizes the role of transverse gradients and the resulting plasma expulsion from the beam path. \AVA{The wave also has a Gaussian temporal profile with a duration of roughly $37\lambda_0/c$, chosen to make the 3D simulation manageable. The Gaussian profile provides a gradual rise and fall, with no separate rise-time parameter.} The plasma is uniform following a short density ramp, with density $n_0=10 n_{cr}$ for each species in the uniform region $(x \ge 0)$.  Additional details of the simulation are provided in \cref{sec:setup}. 

\Cref{fig:density} shows a snapshot of the electron density in three orthogonal slices at $t = 130 \tau$. Additionally, the grayscale overlay shows the envelope of $E_y$ computed as explained in \Cref{app:field-overlay}. \AVA{The positron density is not shown, as it is nearly indistinguishable from the electron density.} \Cref{fig:density} illustrates that the wave does not propagate into the plasma. The result also highlights a key difference between a pair-plasma and an electron-proton plasma. An electron-proton plasma with the same electron density ($n_0 = 10 n_{cr}$) would become relativistically transparent and allow the wave to propagate. The lack of propagation into the pair plasma is consistent with the plane-wave electron density threshold given by \cref{eq:1D_threshold}, as the density in our simulation is well above this threshold.

\begin{figure}
    \centering
    \includegraphics[width=1.0\linewidth]{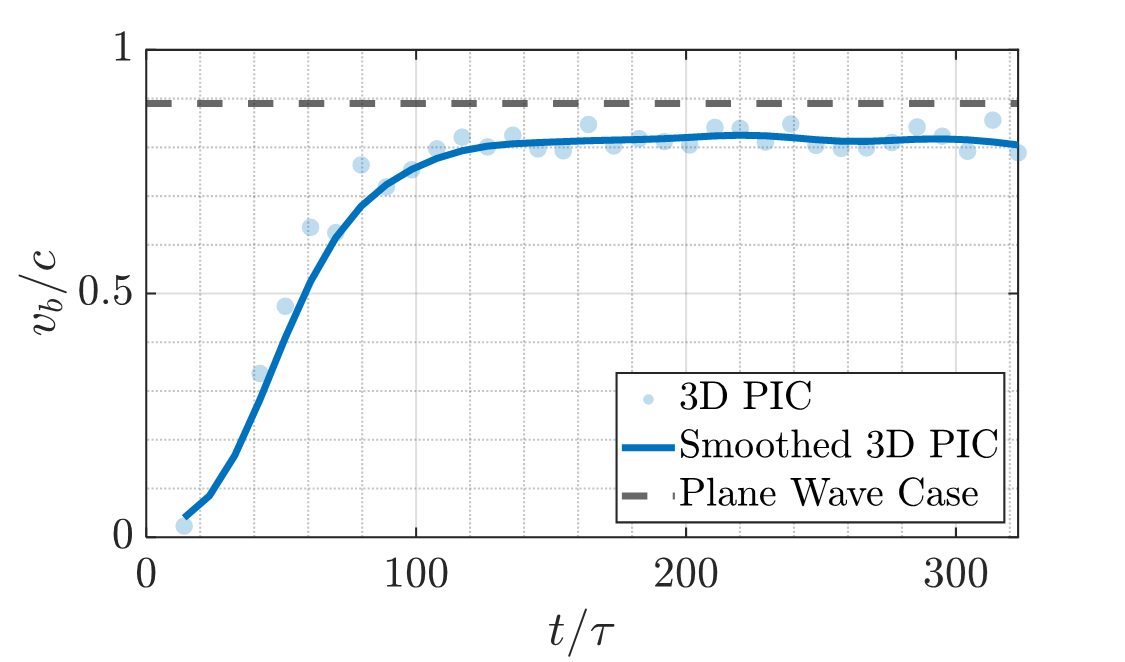}
    \caption{Velocity of the leading boundary of the channel obtained from the 3D PIC simulation. The circular markers show the velocity estimated from the change in the leading-boundary position over intervals of approximately $\tau$. The blue curve is a smoothed result obtained by applying a moving Gaussian average to the marker values. The horizontal dashed line ($v_b/c = 0.89$) shows the velocity of the reflection front predicted by \cref{eq:beta_b_channel} for a linearly polarized plane wave with $a_0=50$ and $n_0=10 n_{cr}$.}
    \label{fig:vb}
\end{figure}

Instead of propagating into the plasma, the wave expels the plasma out of its way, producing a fully evacuated channel inside the plasma. The central part of the beam pushes the plasma forward and forms a dense plug at the leading edge. This is visible in \cref{fig:density}(d), which shows a zoomed-in view of the leading-edge structure. The peak density in the plug is comparable to the estimate from \cref{eq:plug_density}, which gives $n_{\rm{plug}} \approx n_0\psi^{-1/2}\approx 79 n_{cr}$ for $n_0=10 n_{cr}$ and $a_0=50$. However, in contrast to the plane-wave case, the wave does not push all the plasma forward as a piston. The particles near the periphery of the beam are expelled transversely. This produces a visible density ridge along the channel boundary, where the density reaches roughly twice the initial value.

As the wave advances by expelling plasma from its path, the evacuated channel becomes longer. Since the pulse has a finite temporal amplitude profile, the velocity $v_b$ of the leading boundary of the channel changes in time, as shown in \cref{fig:vb}. It is slower in the beginning, before the high-amplitude part of the beam reaches the leading boundary. Once this happens, the velocity remains approximately constant, with $v_b/c \approx 0.83$. This velocity is comparable to the velocity of the reflection front that would be created by a plane wave and is given by \cref{eq:beta_b_channel}. Indeed, for $n_0=10 n_{cr}$ and $a_0=50$, \cref{eq:beta_b_channel} gives $v_b/c \approx 0.89$.

The structure created in the plasma by the beam changes the propagation of the beam itself. It causes the beam to self-focus and provides guiding that prevents the rapid diffraction expected for such a tightly focused beam. In vacuum, the considered beam has a Rayleigh range $x_R = \pi (5\lambda_0)^2/\lambda_0 \approx 79 \lambda_0$. This sets the natural diffraction length. As seen in \cref{fig:density}, there is no appreciable diffraction even though the leading boundary of the channel has advanced $60 \lambda_0$ into the plasma and away from the vacuum focal plane.

Additional snapshots, not shown here, indicate that the evacuated channel does not refill after the wave has passed. In an electron-ion plasma, the charge-separation electric field created during the expulsion of electrons would cause a channel in electron density to refill after the wave passes. No such field is generated in our simulation because electrons and positrons are expelled together, so the channel persists after the wave has passed.

In summary, a wave of finite transverse width creates a fully evacuated channel by expelling some plasma to the side and piling up plasma at the front into a dense plug. The plug, located at the leading boundary of the channel, moves forward with relativistic velocity.

\section{Estimates for energy losses during channel clearing} \label{sec:estimates}

In \cref{sec:channel_formation}, we showed that a relativistically strong electromagnetic wave of finite transverse width expels pair plasma from its path, producing an evacuated channel. The plasma expulsion itself requires energy and is therefore an energy-loss mechanism for the wave. The goal of this section is to provide qualitative estimates for energy losses during channel clearing that can then be used to understand the 3D PIC simulation.

We start with a simple model in which the wave sweeps up all the plasma in its path as a piston. The first step is to find the velocity of the swept-up particles. We use ``sw'' as a subscript to label quantities associated with these particles. In this model, the leading boundary can be treated as the reflection front described in \cref{sec:app:rel-hole-boring-lp}. The front moves with the average velocity $v_b$ given by \cref{eq:beta_b_channel}. In the frame moving with the front, we have an incoming plasma flow that is elastically reflected at the boundary. As a result, the particle velocity changes from $-v_b$ to $v_b$. We find the velocity of the swept-up particles by transforming the velocity after the reflection back to the laboratory frame. We have
\begin{equation}
v_{\rm sw}/c = \beta_{\rm sw} = \frac{2\beta_b}{1+\beta_b^2},
\end{equation}
where $\beta_b$ is given by \cref{eq:beta_b_channel}.

We can now find the kinetic energy of the swept-up particles. Their relativistic factor is
\begin{equation} \label{eq:gamma_sw}
\gamma_{\rm sw} = \frac{1}{\sqrt{1 - \beta_{\rm sw}^2}}
= \frac{1+\beta_b^2}{1-\beta_b^2}
= 2\gamma_b^2 - 1,
\end{equation}
where $\gamma_b$ is the relativistic factor of the reflection front given by \cref{gamma_b}. We are interested in the regime where the swept-up particles are relativistic, which corresponds to $\psi \ll 1$. In this regime, we have $\gamma_b \gg 1$, so
\begin{equation}
    \gamma_{\rm sw} \approx 2\gamma_b^2 \approx \psi^{-1/2}.
\end{equation}
The kinetic energy gained by each swept-up particle is
\begin{equation} \label{energy_sw}
    \varepsilon_{\rm sw} = (\gamma_{\rm sw} - 1)mc^2 .
\end{equation}
The pair plasma has two species with density $n_0$ each, so the energy transferred to the swept-up plasma per unit swept volume is
\begin{equation}
U_{\rm sw}
= 2 n_0 (\gamma_{\rm sw} - 1)mc^2 .
\label{eq:Usw}
\end{equation}
Using \cref{eq:gamma_sw} and assuming $\psi \ll 1$, we obtain
\begin{equation}
U_{\rm sw} \approx 2 n_0 mc^2 \psi^{-1/2}.
\label{eq:Usw_psi}
\end{equation}
This estimate gives the energy cost of sweeping up the plasma.

It is instructive to compare this energy density with the energy density of the incident wave. For a linearly polarized wave in vacuum, the cycle-averaged energy density is
\begin{equation}
U_{\rm wave}
= \frac{E_0^2}{8\pi}
=
\frac{1}{2}a_0^2 n_{cr}mc^2 .
\label{eq:Uwave}
\end{equation}
Using \cref{eq:Usw_psi,eq:Uwave}, we find that
\begin{equation}
U_{\rm sw} \left/ U_{\rm wave} \right. \approx \psi^{1/2},
\label{eq:Usw_Uwave}
\end{equation}
where we again assumed that $\psi \ll 1$.

A large fraction of the wave energy reaching the reflection front is deposited into the plasma, which might at first glance seem to contradict \cref{eq:Usw_Uwave}. During a time interval $\Delta t$, the front sweeps through the plasma over a distance $v_b\Delta t$. Over the same interval, the segment of the incident wave that reaches the moving front has a longitudinal length $(c-v_b)\Delta t$. The ratio of the corresponding volumes is therefore
\begin{equation}
\frac{c-v_b}{v_b} = \frac{1-\beta_b}{\beta_b} = \psi^{1/2},
\end{equation}
where the last equality follows from \cref{eq:beta_b_channel}. Over the same time interval, the energy gained by the swept-up plasma is proportional to $U_{\rm sw}v_b\Delta t$, while the wave energy that reaches the moving front is proportional to $U_{\rm wave}(c-v_b)\Delta t$. Their ratio is therefore
\begin{equation}
\frac{U_{\rm sw}v_b}{U_{\rm wave}(c-v_b)} = \frac{U_{\rm sw}}{U_{\rm wave}}
\frac{\beta_b}{1-\beta_b} \approx 1,
\end{equation}
where we used \cref{eq:Usw_Uwave} and $(1-\beta_b)/\beta_b=\psi^{1/2}$. Thus, an order-unity fraction of the wave energy that reaches the front is transferred to the swept-up plasma, while the remaining fraction is reflected.

A finite-width wave can also expel particles sideways, in the transverse direction. This type of transverse expulsion is common in high-intensity laser-plasma interactions involving electron-ion plasmas because the charge-separation electric field tends to prevent the electron sweep-up discussed above. Motivated by this, we now consider a different model where the pair plasma is expelled only in the transverse direction.

To assess the expulsion, we use a ponderomotive-force estimate and neglect the longitudinal motion of particles. Although the interaction may be too rapid for a strictly cycle-averaged description, this approach provides a useful order-of-magnitude estimate for the transverse expulsion. The ponderomotive force pushes particles away from regions of large field amplitude. For $a \gg 1$, the relativistic ponderomotive potential is of order $a mc^2$, so the amplitude of the ponderomotive force can be roughly estimated as $F_{p} \sim mc^2 |\nabla a|$, where $a$ is the local normalized wave amplitude. The maximum kinetic energy gained during the expulsion is then of order
\begin{equation} \label{energy_perp}
    \varepsilon_{\perp} \approx a_0 m c^2.
\end{equation}
Since a pair plasma has two species with density $n_0$ each, the energy transferred to the plasma per unit evacuated volume is
\begin{equation}
U_{\perp} \approx 2 \kappa n_0 a_0 m c^2.
\label{eq:Uperp}
\end{equation}
The dimensionless coefficient $\kappa$ accounts for the difference between the maximum single-particle estimate and the average energy of particles removed from the evacuated volume.

In the actual simulation, we have a mix of the two mechanisms. Some particles are swept up by the leading boundary, whereas others are expelled in the transverse direction. The scalings for the two mechanisms are different, so, in general, they produce particles with different energies. The ratio of the energies given by \cref{energy_sw,energy_perp} is
\begin{equation}
\varepsilon_{\perp} \left/ \varepsilon_{\rm sw} \right. \approx a_0\sqrt{\psi} = \sqrt{4 n_0 / n_{cr}}.
\end{equation}
This shows that the normalized density, $n_0/n_{cr}$, controls the relative energy scale of the two mechanisms.

In a finite-width wave, the central part of the channel is cleared by a piston-like push, whereas the rest of the channel is cleared by transverse expulsion. This geometric aspect affects the energy partitioning between the two mechanisms. In addition, because transverse expulsion takes place at the periphery, the normalized field amplitude that sets the energy gain is reduced from $a_0$. To account for these aspects, we introduce
\begin{equation}
R_{\rm sw} \equiv r_{\rm sw} / r_{\rm ch},
\end{equation}
where $r_{\rm sw}$ is the radius of the piston-like region and $r_{\rm ch}$ is the radius of cleared channel. \AVA{The value of $R_{\rm sw}$ is determined by the interaction dynamics and can therefore depend on wave parameters such as the pulse duration and transverse width. We do not attempt to estimate this dependence here and instead treat $R_{\rm sw}$ as an input parameter of the simplified energy model (it is not an input parameter of the PIC simulation).}

We can now estimate the energy cost associated with plasma expulsion by both mechanisms. The fraction of the cleared cross-sectional area associated with the piston-like push is approximately $R_{\rm sw}^2$, while the remaining fraction, $1-R_{\rm sw}^2$, is associated with transverse expulsion. The energy spent on clearing the channel per unit evacuated volume can then be estimated as
\begin{equation} \label{U_clear}
    U_{\rm clear} \approx R_{\rm sw}^2 U_{\rm sw}
+ (1-R_{\rm sw}^2) U_{\perp},
\end{equation}
where $U_{\rm sw}$ is the energy transferred to the swept-up plasma per unit swept volume, given by \cref{eq:Usw_psi}, and $U_{\perp}$ is the energy transferred to the transversely expelled plasma per unit evacuated volume, given by \cref{eq:Uperp}. \Cref{eq:Uperp} includes the coefficient $\kappa$ that we use to account for the reduction of the normalized field amplitude $a$ from $a_0$. We set $a = \kappa a_0$ and approximate $a$ as a linearly decreasing function of radius, which then gives $\kappa \approx 1-R_{\rm sw}$. Using this estimate for $\kappa$ in \cref{eq:Uperp}, and then substituting the expressions for $U_{\rm sw}$ and $U_{\perp}$ into \cref{U_clear}, we obtain
\begin{equation} \label{U_clear_2}
    U_{\rm clear} \approx R_{\rm sw}^2 U_{\rm sw} \left[ 1 + 2 (1-R_{\rm sw}) \frac{1-R_{\rm sw}^2}{R_{\rm sw}^2} \sqrt{\frac{n_0}{n_{cr}}} \right].
\end{equation}
The second term in the brackets is the relative contribution of transverse expulsion to the channel-clearing cost.

\section{Wave depletion and particle energetics}
\label{sec:energetics}

In \cref{sec:channel_formation}, we presented a self-consistent kinetic 3D PIC simulation of channel formation in a pair plasma. Motivated by the estimates in \cref{sec:estimates}, we now examine particle energetics in this simulation.

\begin{figure}[tb]
\centering
\includegraphics[width=1.0\linewidth]{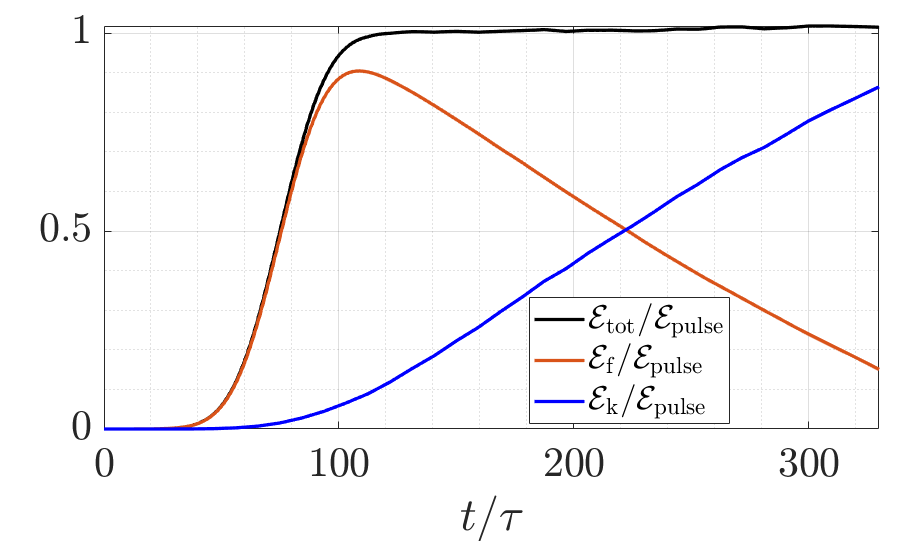}
\caption{Temporal evolution, in the 3D PIC simulation, of the field energy $\mathcal{E}_{\mathrm{f}}$ (red), particle kinetic energy $\mathcal{E}_{\mathrm{k}}$ (blue), and total energy $\mathcal{E}_{\mathrm{tot}}=\mathcal{E}_{\mathrm{f}}+\mathcal{E}_{\mathrm{k}}$ (black), normalized to the full energy in the incident wave $\mathcal{E}_{\mathrm{pulse}}$.}
\label{fig:energy_accounting}
\end{figure}

The estimates in \cref{sec:estimates} suggest weak depletion, meaning that the electromagnetic pulse should propagate a relatively long distance past the initial plasma boundary before losing most of its energy. Capturing this process in a fixed simulation box would require a prohibitively long domain. This creates a computational challenge for a 3D PIC simulation, which we address by using a moving window. The window starts moving after the electromagnetic pulse enters the simulation box. It moves in the direction of wave propagation at the speed of light, $c$, and is sufficiently long that it does not clip the pulse.

While the moving window helps reduce the simulation cost, it complicates the energy accounting. The plasma leaves the moving window through the left boundary and is replaced by cold uniform plasma entering through the right boundary. Additionally, energetic particles leave the window through the transverse boundaries as well. In our case, most energetic particles leave the simulation domain that way. We record the kinetic energy of plasma leaving the window through the left and transverse boundaries and include it in the particle energy accounting. In contrast, the field energy is recorded only inside the moving window. This has little impact on the analysis because the fields are predominantly associated with the wave, which remains inside the moving window for the duration of the simulation.

\Cref{fig:energy_accounting} shows the time evolution of the total kinetic energy of the particles, $\mathcal{E}_{\rm k}$. It also shows the total field energy inside the moving window, $\mathcal{E}_{\rm f}$, and the sum of the two: $\mathcal{E}_{\rm tot} = \mathcal{E}_{\rm f} + \mathcal{E}_{\rm k}$. We normalize the energies to the full energy in the incident wave, $\mathcal{E}_{\rm pulse}$. We use the subscript ``pulse'' to emphasize that the wave has a finite temporal duration and is essentially a pulse. Since the pulse is injected into the simulation box, $\mathcal{E}_{\rm f}$ and $\mathcal{E}_{\rm tot}$ initially increase. After the entire pulse enters the box, $\mathcal{E}_{\rm tot}$ remains essentially constant. Although the pulse reflection takes place, the near constancy of $\mathcal{E}_{\rm tot}$ indicates that the reflected wave carries relatively little energy out of the box. Instead, the field energy is gradually transferred to electrons and positrons. Over the duration of the simulation, about $85\%$ of $\mathcal{E}_{\rm pulse}$ is transferred to particles, with the energy split equally between both species. \AVA{An additional simulation with circular polarization, presented in \cref{app:circular-polarization}, shows that this result is not sensitive to the polarization of the wave.}

Our next step is to identify which particles carry most of the energy transferred from the wave to the plasma. \AVA{We present the analysis for electrons only because the positron results are essentially the same.} A conventional PIC output provides an energy spectrum for each species. However, such a spectrum can be dominated by a large number of low-energy particles, even though they carry relatively little energy. To examine the energy content in the particle population, we focus instead on an energy-weighted spectrum, similarly to what was done in Ref.~[\onlinecite{Tangtartharakul.NJP.2025.Collimated}].

\begin{figure}[b]
    \centering
\includegraphics[width=1.0\linewidth]{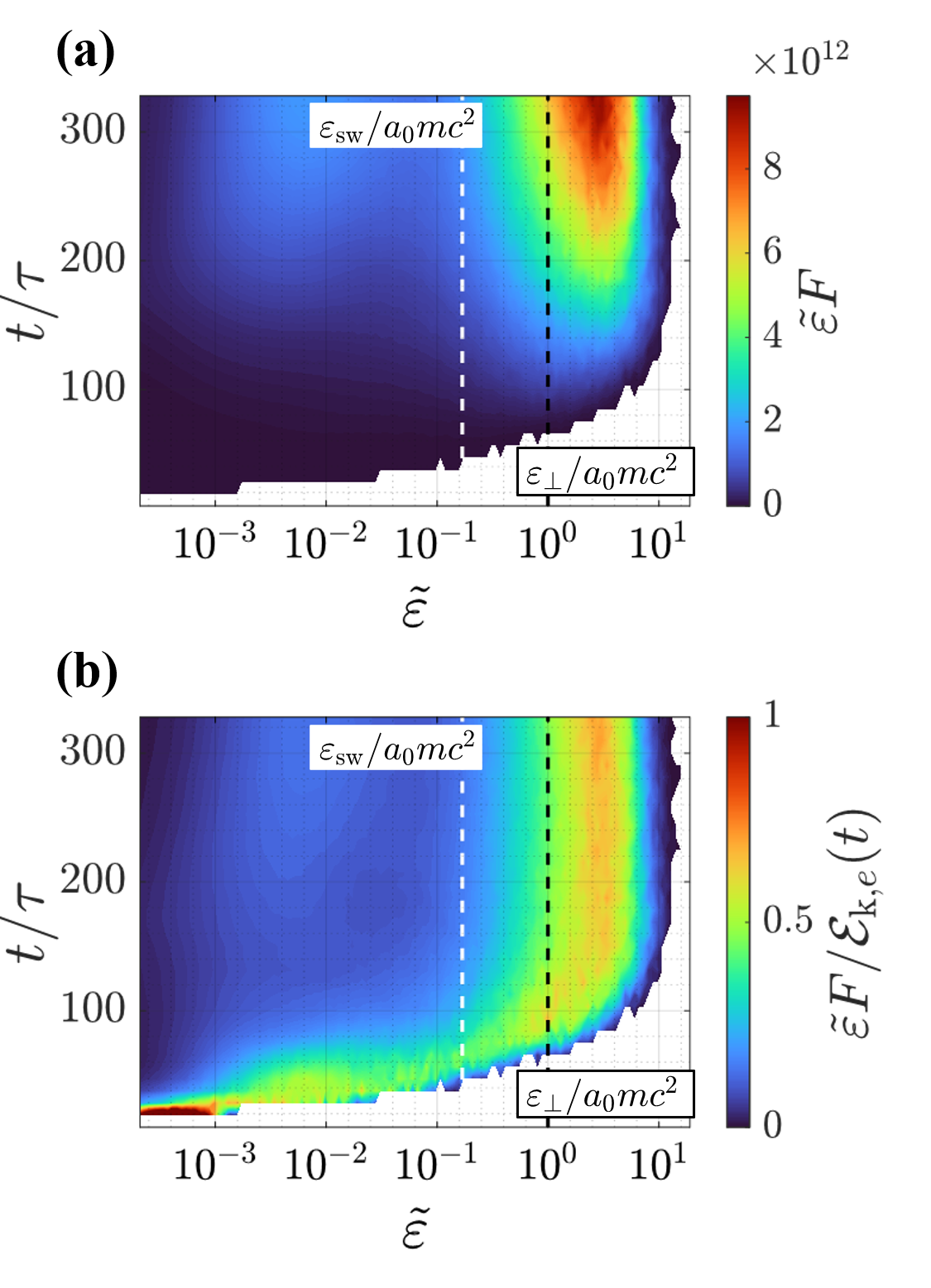}
    \caption{(a) Time evolution of the electron energy-weighted distribution $\tilde{\varepsilon}F$ defined by \cref{EDF-def}. (b) The same distribution normalized at each time by the total kinetic energy of electrons $\mathcal{E}_{\mathrm{k},e}(t)$. In both panels, $\varepsilon_{\rm sw}$ and $\varepsilon_{\perp}$ are the quantities given by \cref{energy_perp,energy_sw} and computed using $a_0=50$ and $n_0=10n_{cr}$. }
    \label{fig:energy_vs_time}
\end{figure}

Before presenting the results, we explain how we define the energy-weighted spectrum. To capture the dynamic range of particle energies, we use a logarithmic scale. We normalize particle kinetic energies $\varepsilon$ to $a_0 mc^2$, which represents a natural scale for particles oscillating in a relativistically strong electromagnetic wave. The normalized energy is
\begin{equation} \label{energy_norm}
    \tilde{\varepsilon} = \varepsilon / a_0 mc^2.
\end{equation}
Using the conventional distribution function $f(\varepsilon) = dN / d\varepsilon$, where $N$ is the particle number, we define
\begin{equation} \label{EDF-def}
    F = dN / d \log_{10} \tilde{\varepsilon} = \ln(10) \varepsilon f(\varepsilon).
\end{equation}
The energy-weighted distribution is then $\tilde{\varepsilon} F$. Its integral over $\log_{10}\tilde{\varepsilon}$ gives the total kinetic energy for the considered species normalized to $a_0mc^2$.

\begin{figure}[tb]
    \centering
    \includegraphics[width=1.0\linewidth]{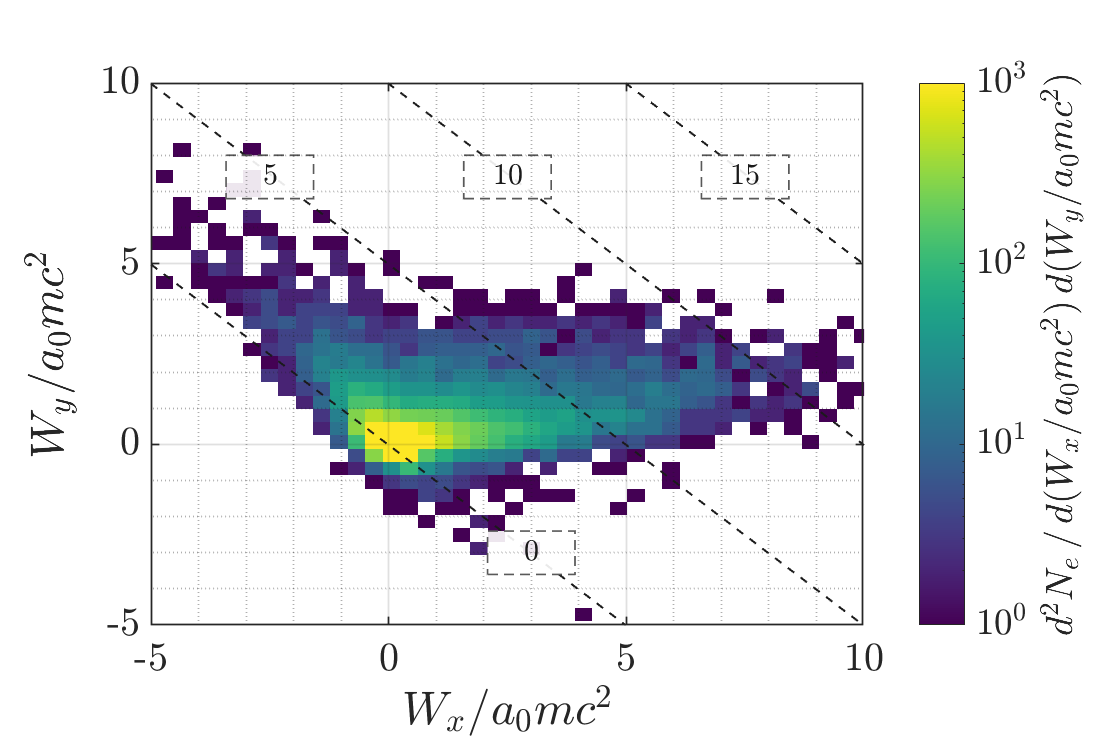}
    \caption{Two-dimensional histogram of macro-particles representing electrons ($1/200$ random sample) showing accumulated work components $(W_x,W_y)/a_0mc^2$ from $t/\tau=0$ to $t/\tau=185$. The color scale is normalized so that the integral over the plot gives the total electron number $N_e$. Dashed diagonal lines indicate constant $(W_x+W_y)/a_0mc^2$, labeled in increments of 5.}
    \label{fig:work_xy}
\end{figure}

\Cref{fig:energy_vs_time}(a) shows the time evolution of the energy-weighted distribution $\tilde{\varepsilon}F$ for the electrons. The distribution includes particles inside the moving window and particles that have left it. The vertical dashed lines show two energy estimates from \cref{sec:estimates}. The white dashed line is the characteristic energy of swept-up particles, $\varepsilon_{\rm sw}$, given by \cref{energy_sw}. The black dashed line is the estimate for the characteristic kinetic energy associated with transverse expulsion, $\varepsilon_{\perp}$, given by \cref{energy_perp}. For our set of parameters, $\varepsilon_{\rm sw} < \varepsilon_{\perp}$. The feature that clearly stands out is that most of the energy is carried by particles with energies well above $\varepsilon_{\perp}$. This places the dominant energy-carrying population well above the characteristic energies expected from channel clearing.

\Cref{fig:energy_vs_time}(b) presents the same distribution, but normalized at each time by the total kinetic energy of electrons, $\mathcal{E}_{\mathrm{k},e}(t)$. Note that $\mathcal{E}_{\mathrm{k},e}(t)$ is approximately half of the total particle kinetic energy, $\mathcal{E}_{\mathrm{k}}(t)$, whose time evolution is shown in \cref{fig:energy_accounting}. The purpose of the normalization is to assess the relative weight of the energetic particles as the wave continues to lose energy. The normalized distribution indicates that once the population of particles with $\varepsilon > \varepsilon_{\perp}$ is generated by about $t \approx 100 \tau$, its relative weight remains roughly unchanged. The shape of the distribution for $\varepsilon > \varepsilon_{\perp}$ also remains relatively unchanged. Together with the continued wave depletion shown in \cref{fig:energy_accounting}, this suggests continued production of energetic particles.

The analysis presented here shows that channel clearing alone cannot account for the energy loss experienced by the wave in our setup. The energy accounting reveals that there is another particle-acceleration mechanism that generates the most energetic populations \AVA{of both electrons and positrons}. The generation of \AVA{these} population is responsible for the relatively rapid depletion seen in \cref{fig:energy_accounting}, where the pulse, whose duration is about $75\tau$, loses most of its energy by $t \approx 350\tau$.

\begin{figure}[t]
    \centering
    \includegraphics[width=0.9\linewidth]{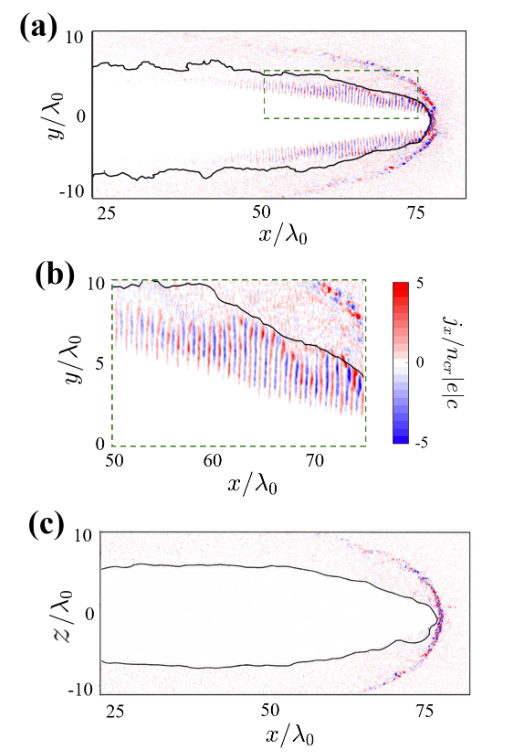}
    \caption{Longitudinal plasma current density $j_x$ at $t/\tau=148$. Panels (a) and (b) show the polarization plane, the $(x,y)$-plane at $z=0$, with panel (b) showing a zoomed-in view of the boxed region in panel (a). Panel (c) shows the orthogonal plane, the $(x,z)$-plane at $y=0$. The color scale gives $j_x/(n_{cr}|e|c)$. The black contour marks where the smoothed electron density reaches $n_e=n_{cr}$, which we use as a marker for the approximate boundary of the evacuated channel.}
    \label{fig:plasma_current}
\end{figure}

\section{Particle acceleration inside the channel}
\label{sec:channel_acc}

\Cref{sec:energetics} shows that the most energetic particles are responsible for the rapid wave depletion. We now identify how these particles gain their energy.

To determine how the energetic particles gain their energy, we examine the work done by the electric field on each particle. The instantaneous rate of energy transfer to a particle with charge $q$ is $q (\bm{E} \cdot \bm{v})$. The accumulated work, which gives the cumulative energy transferred by the electric field, is therefore
\begin{equation}
W(t) = \int_0^t q \left( \bm{E}\cdot\bm{v} \right) dt'.
\end{equation}
To identify which field component is responsible for the energy transfer, we decompose this work into contributions from the three electric-field components:
\begin{equation}
W_{\alpha}(t) = \int_0^t q E_{\alpha} v_{\alpha} dt',
\quad \alpha = x,y,z.
\end{equation}
The total accumulated work is then
\begin{equation}
W(t) = W_x(t) + W_y(t) + W_z(t).
\end{equation}

Before applying this diagnostic, it is useful to recall the electric field structure of the incident linearly polarized wave in vacuum. The dominant component is $E_y$. In the case of a plane wave, it would be the only component. The incident wave has a finite transverse width, so it also has a longitudinal electric field in order to satisfy $\nabla \cdot \bm{E} = 0$. The magnitude of the longitudinal field is small if the local transverse width of the wave, $R$, exceeds the vacuum wavelength $\lambda_0$. It can be estimated as $|E_x| \approx |E_y| \lambda_0 / R$. Guided by these considerations, we focus on $W_y$ and $W_x$. Particle tracking confirms that $W_z$ is negligible.

\begin{figure}[t]
    \centering
    \includegraphics[width=0.9\linewidth]{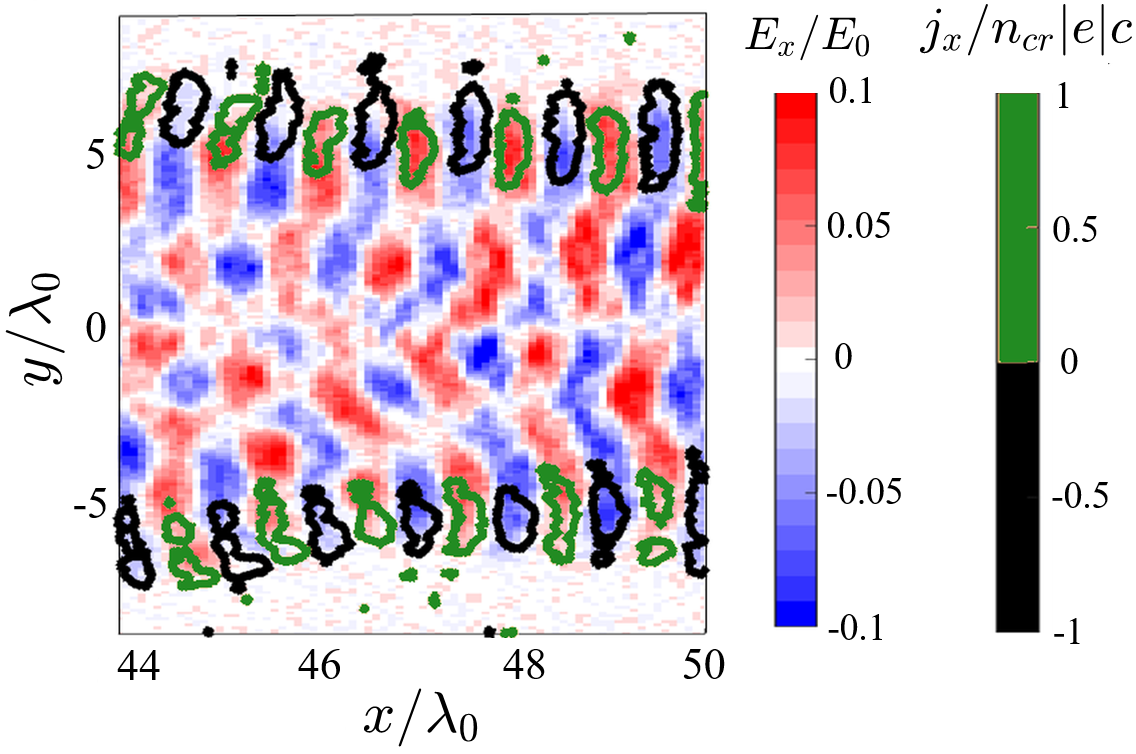}
    \caption{Longitudinal electric field $E_x$ in the $(x,y)$-plane at $z=0$ and $t/\tau=111$. The black and green contours mark the longitudinal current-density levels $j_x=\pm |e|n_{cr}c$.}
    \label{fig:current_and_Ex}
\end{figure}

\begin{figure*}[!htb]
    \centering
    \includegraphics[width=0.8\linewidth]{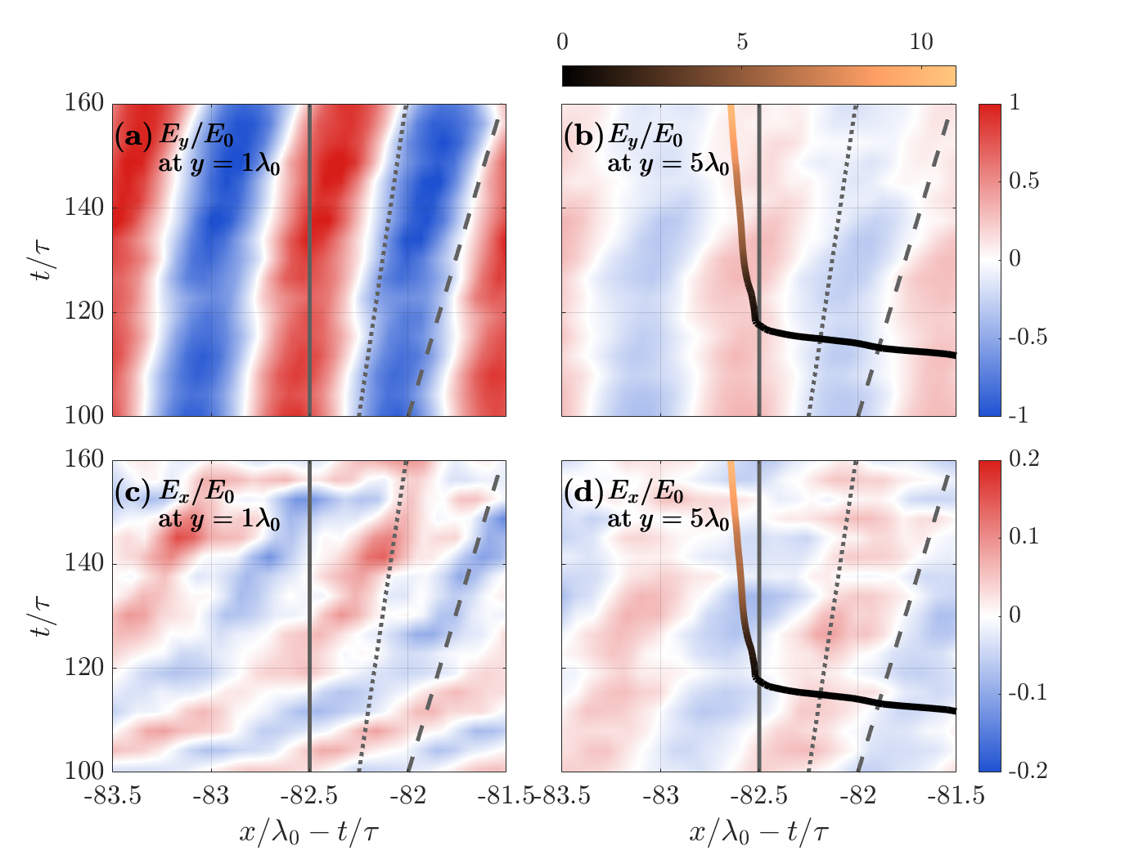}
    \caption{Time evolution of the longitudinal position of the tracked electron (b,d) and field lineouts for $E_y$ (a,b) and $E_x$ (c,d) at two transverse positions. The color along the electron trajectory indicates the normalized kinetic energy $\tilde{\varepsilon}=\varepsilon/(a_0mc^2)$. The solid, dotted, and dashed gray lines show reference motion of wave fronts with velocities $c$, $1.004c$, and $1.008c$, respectively. Panels (a) and (c) show the field lineouts at $y=\lambda_0$ and $z=0$, whereas panels (b) and (d) show the lineouts at $y=5\lambda_0$ and $z=0$, close to the transverse location of the tracked electron.}
    \label{fig:Ex_phase}
\end{figure*}

\Cref{fig:work_xy} shows a histogram separating $W_x$ and $W_y$ for electrons in the simulation at $t = 185\tau$. The color shows the electron distribution over $W_x$ and $W_y$, normalized such that the integral over the entire plot gives the total electron number $N_e$. The histogram is computed from a random subsample containing $1/200$ of the macro-particles representing electrons. The dashed diagonal lines indicate constant $W_x + W_y$ and are essentially lines of constant $\gamma$ in our setup, since $W_z$ is negligible. The number next to each line is $(W_x + W_y)/a_0mc^2$, making it easier to correlate this distribution with that shown in \cref{fig:energy_vs_time}. The histogram reveals that the highest-energy electrons have $W_x \gg W_y$. This means that the work performed by the longitudinal electric field is the dominant contribution to the energy gain of these particles. \AVA{A similar analysis for positrons yields the same conclusion.}

To identify where longitudinal energy transfer can occur, we examine the longitudinal plasma current density $j_x$, since the local power density associated with the longitudinal field is $j_xE_x$. To determine whether the current structure depends on the orientation of the incident-wave electric field relative to the channel wall, we compare $j_x$ in two orthogonal planes that both pass through the channel axis. The first is the $(x,y)$-plane at $z=0$, which we refer to as the polarization plane because it is aligned with the electric-field polarization of the incident wave. The second is the $(x,z)$-plane at $y=0$, which we refer to as the orthogonal plane because the electric-field polarization of the incident wave is perpendicular to this plane.

\begin{figure}[b]
    \centering
    \includegraphics[width=1.0\linewidth]{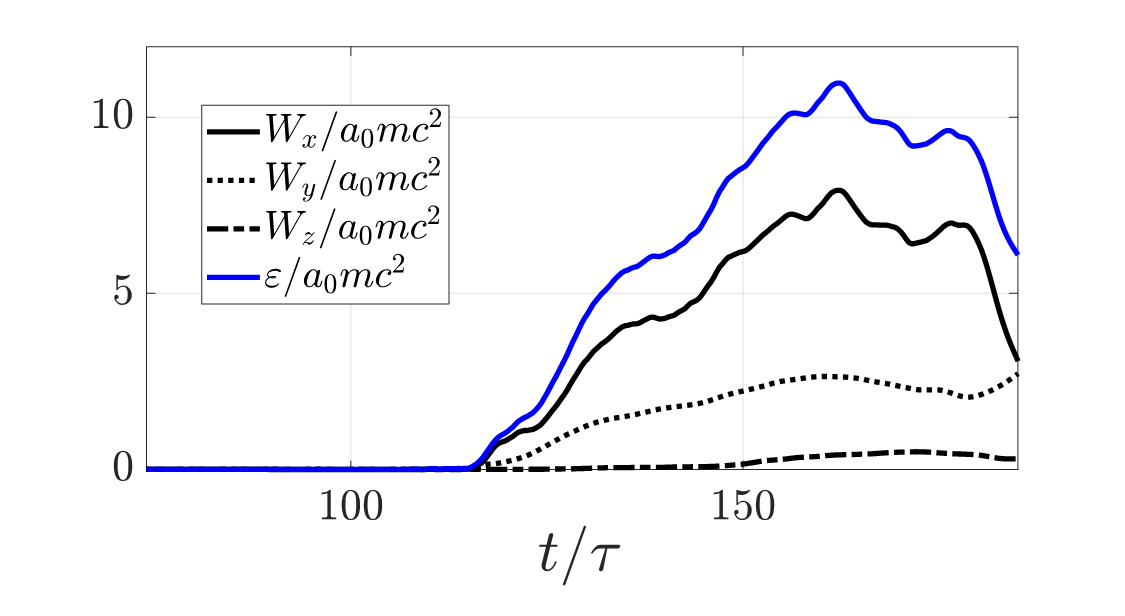}
    \caption{Accumulated work from different electric-field components and kinetic energy for the electron tracked in \cref{fig:Ex_phase}.}
    \label{fig:work_traj}
\end{figure}

\Cref{fig:plasma_current}(a) shows a snapshot of $j_x$ in the polarization plane at $t = 148\tau$. The black contour marks where the smoothed electron density reaches $n_e=n_{cr}$, which we use as a marker for the approximate boundary of the evacuated channel. \AVA{The key feature is the appearance of strong longitudinal current density near the channel walls, while the central region carries little current because the channel is almost completely evacuated.} The current density has a clear alternating structure, seen in both the full view in panel~(a) and the zoomed-in panel~(b). 

\AVA{\Cref{fig:plasma_current}(c) shows $j_x$ in the orthogonal plane at the same time. The alternating current-density structure is clearly absent near the channel wall. In contrast, an additional simulation with circular polarization, presented in \cref{app:circular-polarization}, shows strong longitudinal current in both planes, including the $(x,z)$-plane shown here. This difference points to a direct connection between current generation and the transverse electric field. A strong electric-field component normal to the channel boundary is needed to inject particles into the channel where they can subsequently gain energy from the longitudinal electric field. For a linearly polarized wave, particle injection and the resulting longitudinal current are localized near the polarization plane. For a circularly polarized wave, the transverse electric field rotates, so particle injection and the resulting longitudinal current are distributed around the entire periphery of the channel. We therefore focus on the polarization plane in the following analysis of the linearly polarized case.}

\Cref{fig:current_and_Ex} shows a profile of $E_x$ in the polarization plane. The figure also shows two contours of the longitudinal current density, $j_x = \pm |e| n_{cr} c$. The longitudinal electric field remains strong where the current bunches are located. The longitudinal period of $E_x$ matches that of the $j_x$ structure. Moreover, the peaks of positive $j_x$ overlap with the peaks of positive $E_x$, while the peaks of negative $j_x$ overlap with the peaks of negative $E_x$. The simulation data shows that the positive peaks of $j_x$ are dominated by forward-moving positrons, whereas the negative peaks of $j_x$ are dominated by forward-moving electrons. This species separation, together with the phase alignment between $j_x$ and $E_x$, explains why the longitudinal electric field transfers energy to both species despite the alternating structure of the current.

The alignment between $j_x$ and $E_x$ persists as the wave propagates forward. We observed this by examining multiple consecutive snapshots. The aligned structures move together for a time much longer than the wave period. This observation suggests that particles can surf with the longitudinal field while gaining energy.

To confirm the surfing process, we tracked an electron picked from a negative current bunch located at $x \approx 65.5\lambda_0$ in \cref{fig:plasma_current}. We also recorded a lineout of $E_x$ along $x$ as a function of time, while keeping $y/\lambda_0 = 5$ and $z/\lambda_0 = 0$ fixed. This transverse position was chosen because it passes through the selected current bunch in \cref{fig:plasma_current}. Since the electron moves primarily forward, this lineout provides a proxy for the longitudinal field sampled by the electron. Together, the particle trajectory and the field lineout allow us to examine the electron's longitudinal position relative to the field structure.

\Cref{fig:Ex_phase}(d) shows how the longitudinal profile of $E_x$ evolves with time in a window moving at the speed of light. The horizontal coordinate in this window is $x/\lambda_0 - t/\tau$. The moving window makes it easier to compare the motion of the electron with the motion of the field structure, since both move forward with velocities close to $c$. In this plot, a vertical line would represent forward motion at the speed of light. The electron's position in the moving window is shown by the color-coded curve, where the color represents the normalized kinetic energy. The electron moves to the left in the window because $v_x < c$, while the alternating pattern of $E_x$ moves to the right. The figure confirms the surfing picture suggested by the current-field alignment: the tracked electron remains in the accelerating phase of the longitudinal electric field, $E_x < 0$, for $\Delta t \approx 50\tau$ while gaining energy.

The accumulated work done by different electric-field components on the tracked electron is given in \cref{fig:work_traj}. Most of the work is indeed done by the longitudinal electric field $E_x$ over an extended period, $\Delta t \approx 50\tau$. The transverse electric field $E_y$ experienced by the tracked electron is generally much stronger than $E_x$, as seen in \cref{fig:Ex_phase}(b), but its accumulated work is several times smaller. This is in part because the electron motion is predominantly longitudinal, which suppresses the work done by the transverse field. Consistent with the earlier statement in this section, the contribution from $E_z$ is negligible. The tracking thus confirms that the energy gain is dominated by work done by the longitudinal electric field as the electron surfs forward with the field structure.

\begin{figure*}
    \centering
    \includegraphics[width=.95\linewidth]{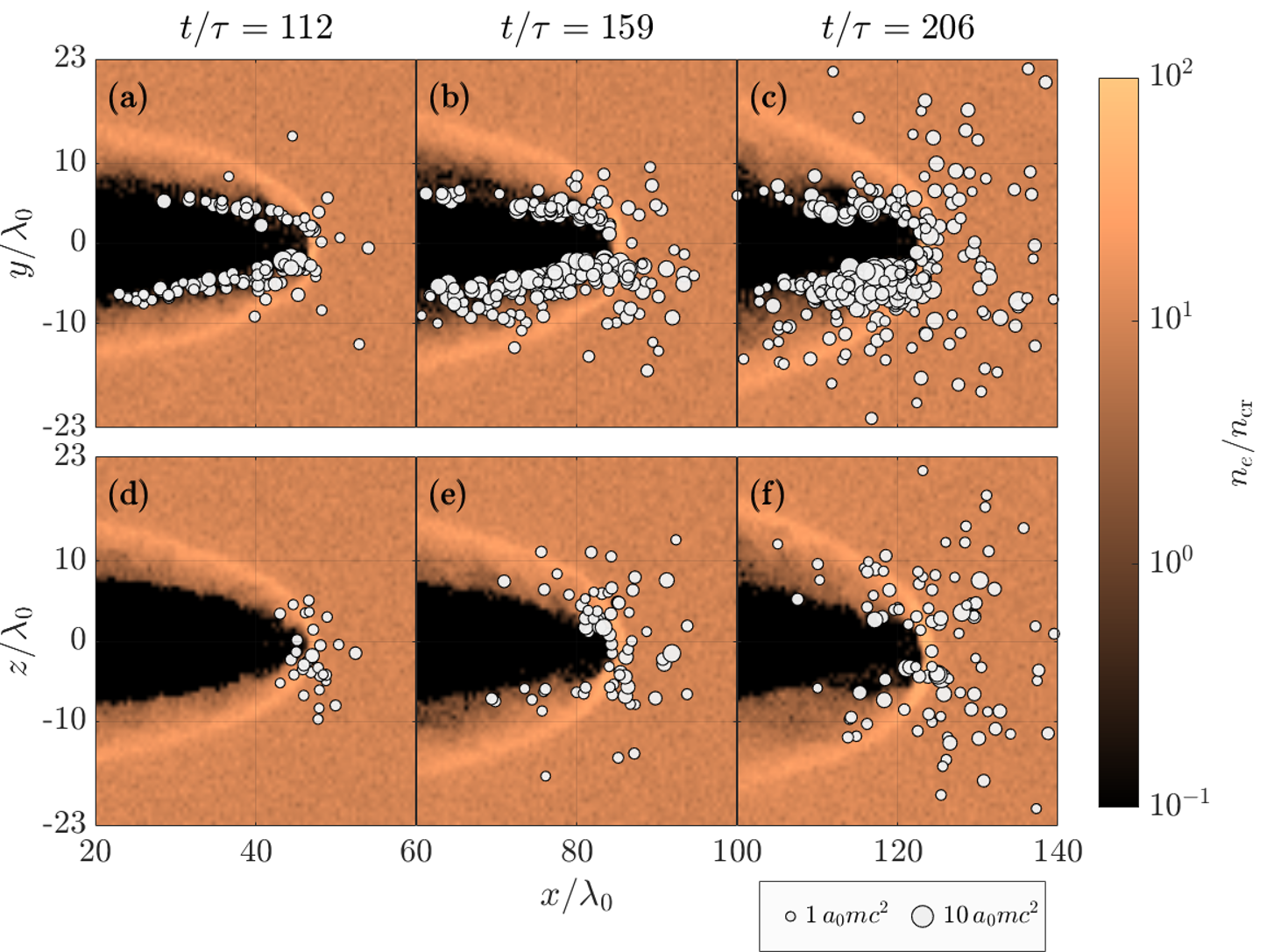}
    \caption{Evolution of the energetic electron population with kinetic energy $\varepsilon>a_0mc^2$. Panels (a)--(c) show the polarization plane at $t/\tau=112$, 159, and 206. Panels (d)--(f) show the orthogonal plane at the same times. In each panel, circles show a random 1-in-200 sample of electrons selected from a slab of thickness $0.2\lambda_0$ centered on the corresponding plane. The marker area scales linearly with $\varepsilon$. For reference, the background shows $n_e/n_{cr}$ in the central slice of each slab.}
    \label{fig:scatter}
\end{figure*}

To conclude this section, \cref{fig:scatter} shows energetic electrons with $\varepsilon>a_0mc^2$ in the polarization and orthogonal planes. In each case, we select electrons from a slab of thickness $0.2\lambda_0$ and display a random 1-in-200 sample. For reference, the background color shows the electron density. In all three snapshots, there is a population of energetic electrons inside the channel near the wall. This supports the surfing picture developed above. The absence of a similar population in the orthogonal plane shows that the scenario plays out primarily in the polarization plane where a strong transverse field can inject particles into the channel.

\AVA{The surfing picture applies to both species. The electron tracked above surfs with $E_x<0$, while positrons surf in the opposite phase, with $E_x>0$. In both cases, the longitudinal electric force accelerates the particles forward.} The key conclusion is that the most energetic particles responsible for the rapid wave depletion gain their energy from the longitudinal electric field $E_x$ while moving forward.

\section{Estimate for energy gain from the longitudinal field}
\label{sec:energy_gain_estimate}

We have shown that energetic electrons and positrons gain most of their energy from the longitudinal field while moving forward along the channel. The purpose of this section is to provide estimates and scalings for the maximum particle energy gain. \AVA{Even though we develop our estimates for an electron, the same estimates apply to positrons, which surf with $E_x>0$ rather than $E_x<0$.}

These estimates leverage the key feature of the electron dynamics: the electron moves primarily forward while gaining energy. This might appear contradictory, as the electric-field plots in \cref{fig:Ex_phase} clearly show that $|E_y| \gg |E_x|$ at the electron location. However, the electron also experiences a transverse magnetic field $B_z$ of nearly equal strength. For a forward-moving electron, the force associated with $B_z$ is opposite to the force associated with $E_y$. This compensation is particularly strong for an ultra-relativistic electron. To show this, we evaluated the transverse component of $\bm{E}+(1/c)[\bm{v} \times \bm{B}]$ while setting $v_\parallel=c$. Its absolute value is an order of magnitude smaller than $|E_y|$. \AVA{Since we use the total fields from the PIC simulation, this force compensation automatically accounts for the fields of the wave and the transverse electric and magnetic fields generated by the relativistically moving particle bunches. The fields generated by these bunches are not negligible, but their electric and magnetic contributions also largely compensate in the transverse force~\cite{Gong.PPCP.2019.Hollow}. This result supports the simplification of treating the electron as moving primarily forward over the energy-gain interval.}

%These estimates leverage the key feature of the electron dynamics: the electron moves primarily forward while gaining energy. This might appear contradictory, as the electric-field plots in \cref{fig:Ex_phase} clearly show that $|E_y| \gg |E_x|$ at the electron location. \rc{For a relativistically forward-moving bunch, the self-generated field is anisotropic: the component parallel to the motion, $E_x$, is suppressed relative to the transverse field by a factor of order $1/\gamma_\parallel$, where $\gamma_\parallel$ is the Lorentz factor associated with the bunch's longitudinal motion along the channel, while the transverse electric and magnetic fields remain substantial~\cite{Gong.PPCP.2019.Hollow}. The relevant transverse quantity is the full Lorentz force, rather than $B_z$ alone; for a forward-moving electron, the force associated with $B_z$ is opposite to that associated with $E_y$.} This compensation is particularly strong for an ultra-relativistic electron. To show this, we evaluated the transverse component of $\bm{E}+(1/c)[\bm{v} \times \bm{B}]$ while setting $v_\parallel=c$. Its absolute value is an order of magnitude smaller than $|E_y|$. This supports our simplification of treating the electron as moving primarily forward over the energy-gain interval.

We can now obtain a basic estimate for the energy gain using a simple model where an electron moves forward in a longitudinal electric field $E_x$. The field pattern moves forward along the $x$-axis with velocity $v_*$, while the electron has longitudinal velocity $v_x$. We assume $v_* > v_x$. If the wavelength of the field oscillations is $\lambda_*$, then the electron can gain energy while remaining in an accelerating field segment of length $\lambda_*/2$. The electron slips backward relative to the field pattern at the speed $v_* - v_x$, so the energy-gain time is approximately
\begin{equation} \label{Delta_t_estimate}
    \Delta t \approx \frac{\lambda_*}{2(v_* - v_x)}.
\end{equation}
If the peak amplitude of the longitudinal field is $E_*$, then the corresponding energy gain is
\begin{equation} \label{eq:gain}
    \Delta \varepsilon \approx |e| E_* v_x \Delta t
    \approx \frac{|e|E_*\lambda_*}{2}\frac{v_x}{v_* - v_x}.
\end{equation}
This estimate shows that the slip speed, $v_* - v_x$, is the key parameter controlling the energy gain.

To make further simplifications, we use the electron tracking example shown in \cref{fig:Ex_phase}. In this case $v_* \approx 1.008 c$ and $v_x \approx 0.998 c$. Since the electron velocity is always less than $c$, we explicitly separate the slip speed into two positive contributions:
\begin{equation}
    v_* - v_x = (v_* - c) + (c - v_x).
\end{equation}
For the tracked electron, $v_* - c$ is larger than $c - v_x$, so we neglect the second term to obtain a simple estimate. Moreover, in the numerator of \cref{eq:gain} we can replace $v_x$ by $c$. As a result, \cref{eq:gain} reduces to
\begin{equation} \label{eq:gain_2}
    \Delta \varepsilon
    \approx \frac{|e|E_*\lambda_*}{2} \frac{c}{v_* - c}.
\end{equation}
It is convenient to introduce a relative degree of superluminality of the field structure,
\begin{equation}
    \delta u = \frac{v_* - c}{c},
\end{equation}
and to normalize the energy gain to $a_0 m c^2$, to obtain
\begin{equation}
    \Delta \tilde{\varepsilon}
    \equiv \frac{\Delta \varepsilon}{a_0 m c^2}
    \approx
    \frac{\pi}{\delta u} \frac{E_*}{E_0}
    \frac{\lambda_*}{\lambda_0}.
    \label{eq:gain_norm}
\end{equation}

We now apply this estimate to the electron tracking example shown in \cref{fig:Ex_phase}(d). The $E_x$ pattern approximately follows the dashed line, so that $v_* \approx 1.008c$. This gives $\delta u \approx 8 \times 10^{-3}$. The peak value of $|E_x|$ in the plotted region is of order $0.1E_0$. Since \cref{eq:gain_norm} uses a constant field amplitude, we take $E_*$ to be the representative field sampled by the electron over the accelerating interval and use $E_*/E_0 \approx 0.03$. The longitudinal wavelength of the field structure is close to $\lambda_0$, so that $\lambda_*/\lambda_0 \approx 1$. Substituting these values into \cref{eq:gain_norm}, we obtain $\Delta \tilde{\varepsilon} \approx 11.3$. This is consistent with the normalized energy gain of the tracked electron in \cref{fig:work_traj}.

Our result shows that the simple model adequately captures the key physics elements despite its simplifications. It is particularly worth noting that the final estimate is not sensitive to $v_x$ for $v_* - c \gg c - v_x$. The implication is that the estimate is also applicable when the particle moves at an angle $\theta$ to the direction of field propagation. For ultra-relativistic electrons, $c-v_x \approx \theta^2 c/2$ when $\theta \gtrsim 1/\gamma$. Then our estimate is applicable if $\theta \ll \sqrt{2\delta u}$. In the tracking example, $v_x/c \approx 0.998$, which corresponds to $\theta \approx 3.6^\circ$. This is smaller than $\sqrt{2\delta u} \approx 7^\circ$, so the use of \cref{eq:gain_norm} is justified.

Our next step is to provide basic estimates for $E_x$, as this is the field that determines the energy gain. However, the dominant field component in the incident wave and then inside the channel is the transverse field $E_y$. Therefore, the estimates must be built by deducing the structure of $E_x$ from $E_y$. We consider two contributions to $E_x$ that arise from different aspects of the channel geometry and estimate each separately. \AVA{It is worth pointing out that the relativistically moving particle bunches can also generate $E_x$. However, this contribution is suppressed by the Lorentz factor associated with their forward motion~\cite{Gong.PPCP.2019.Hollow}, so we neglect it in the following estimates.}

The channel effectively serves as a waveguide. We first consider the simpler case of a straight waveguide. We label the longitudinal electric field and the relative degree of superluminality in this straight-waveguide approximation with a superscript ``str'': $E_x^{\rm str}$ and $\delta u^{\rm str}$. Although the linear theory does not directly apply to our case, it still provides a useful conceptual reference point. In a straight waveguide, a linear electromagnetic wave is a superposition of discrete eigenmodes. We focus on the fundamental mode, since $E_y$ inside the channel does not show pronounced radial oscillations. We denote the longitudinal wavenumber of the mode as
\begin{equation}
    k_* \equiv 2\pi / \lambda_* .
\end{equation}
The transverse confinement of the mode imposes a transverse wavenumber
\begin{equation}
    k_\perp \approx \alpha / r_{\rm ch},
\end{equation}
where $r_{\rm ch}$ is the channel radius and $\alpha$ is an order-unity constant.

\AVA{To find the longitudinal field, we use $\nabla \cdot \bm{E} \approx 0$. This condition requires the transverse derivative of $E_y$ to be balanced by the longitudinal derivative of $E_x$. The latter is simply estimated as $k_* E_x^{\rm str}$. To estimate the former, we approximate the amplitude of the transverse guided field by the incident-wave amplitude $E_0$. Since this field varies over the characteristic transverse scale $k_\perp^{-1}$, we have $\partial E_y/\partial y \sim k_\perp E_0$. The condition $\nabla \cdot \bm{E} \approx 0$ therefore gives}
\begin{equation}
    k_* E_x^{\rm str} \approx k_\perp E_0 .
\end{equation}
It is convenient to introduce
\begin{equation}
    k_0 \equiv \omega_0/c = 2\pi/\lambda_0,
\end{equation}
the vacuum wavenumber for the frequency of the incident wave, and rewrite our result as
\begin{equation} \label{E_x_str}
    \frac{E_x^{\rm str}}{E_0}
    \approx
    \frac{\alpha}{2 \pi} \frac{\lambda_0}{r_{\rm ch}} \frac{k_0}{k_*}.
\end{equation}
This is the amplitude of the longitudinal field in a straight waveguide.

Another quantity needed for the energy-gain estimate is $\delta u$. By definition, $\delta u$ depends on the velocity $v_*$ of the field structure along the channel. In the guided-mode picture, this velocity is the longitudinal phase velocity of the mode, $v_* = v_{\rm ph} = \omega_0 / k_* $. Since $k_0 = \omega_0/c$, we have
\begin{equation} \label{eq:delta_u_estimate}
    \delta u^{\rm str} \approx (v_{\rm ph} - c)/c = (k_0 - k_*)/k_* .
\end{equation}
To estimate the difference between $k_0$ and $k_*$, we treat the field inside the evacuated channel as an electromagnetic wave with transverse wavenumber $k_\perp$, which gives
\begin{equation}
    k_*^2 + k_\perp^2 \approx k_0^2 .
\end{equation}
For $k_\perp \ll k_0$, we then have
\begin{equation}
    k_* \approx k_0 \left( 1 - \frac{1}{2}\frac{k_\perp^2}{k_0^2} \right).
\end{equation}
Using this in \cref{eq:delta_u_estimate}, where we can replace $k_*$ by $k_0$ in the denominator, yields
\begin{equation} \label{delta_u_str}
    \delta u^{\rm str} \approx \frac{1}{2}\frac{k_\perp^2}{k_0^2} \approx \frac{1}{2} \left( \frac{\alpha}{2 \pi} \right)^2 \left( \frac{\lambda_0}{r_{\rm ch}} \right)^2.
\end{equation}

The straight-waveguide estimate assumes a fixed channel radius $r_{\rm ch}$. In our case, the channel is not straight: its radius decreases with $x$. This convergence generates an additional contribution to $E_x$ that the straight-waveguide picture does not capture. We label this contribution to the longitudinal electric field and the corresponding contribution to the relative degree of superluminality with a superscript ``conv'': $E_x^{\rm conv}$ and $\delta u^{\rm conv}$.

The other contribution can be understood as reflection off the channel wall in the polarization plane. To produce a simple estimate, we consider a plane wave with amplitude $E_0$ and phase velocity $c$ propagating along the $x$-axis, with its electric field polarized along $y$. The wave reflects off a wall whose cross-section in the polarization plane is tilted, with the angle between the wall and the axis being $\theta_x$. The incident wave has no longitudinal electric field in this simplified estimate, but the reflection produces a wave with an $x$-component of the electric field. For a specular reflection, the reflected wave propagates at an angle $2\theta_x$ relative to the axis. The longitudinal field component in this wave is $E_x^{\rm conv} \approx E_0 \sin(2\theta_x)$. Its phase velocity along $x$ is $v_* = c/\cos(2\theta_x)$. The angle $\theta_x$ is set by the local slope of the channel wall,
\begin{equation}
    \theta_x \approx \left| \Delta r_{\rm ch} \right| /\Delta x.
\end{equation}
For a small tilt ($\theta_x \ll 1$), we simply have
\begin{equation} \label{E_x_conv}
    E_x^{\rm conv}/E_0 \approx 2\theta_x
\end{equation}
and
\begin{equation} \label{delta_u_conv}
    \delta u^{\rm conv} \approx 2\theta_x^2 .
\end{equation}
Unlike $E_x^{\rm str}$ and $\delta u^{\rm str}$, which depend on the channel radius $r_{\rm ch}$, this contribution depends only on the local wall tilt in the polarization plane.

We next compare the expressions for $E_x^{\rm str}$ and $E_x^{\rm conv}$. It follows from \cref{E_x_str,E_x_conv} that the contribution associated with the reflection becomes important, $E_x^{\rm conv} \gtrsim E_x^{\rm str}$, if
\begin{equation} \label{slope_cond}
    \frac{|\Delta r_{\rm ch}|}{\Delta x}  \gtrsim \frac{\alpha}{4\pi}\frac{\lambda_0}{r_{\rm ch}}.
\end{equation}
This gives a specific condition for the tilt of the channel wall. It can be seen directly from \cref{delta_u_str,delta_u_conv} that if this condition is satisfied, then $\delta u^{\rm conv} \gtrsim \delta u^{\rm str}$. Therefore, we find that the tilt of the channel wall increases the longitudinal field, while, at the same time, increasing the relative degree of superluminality.

Our finding has a direct implication for the energy gain, whose generic estimate is given by \cref{eq:gain_norm}. If the tilt is sufficiently weak, such that $E_x^{\rm str} \gg E_x^{\rm conv}$, then the straight-channel contribution determines the energy gain. Substituting the expressions given by \cref{E_x_str,delta_u_str} into \cref{eq:gain_norm}, we find that $\Delta \tilde{\varepsilon} \approx \sqrt{2}\pi / \sqrt{\delta u^{\rm str}}$. Here we additionally set $\lambda_* \approx \lambda_0$, which is valid for $\delta u^{\rm str} \ll 1$. When the tilt satisfies \cref{slope_cond}, the estimate must instead use the field associated with reflection. Substituting the expressions given by \cref{E_x_conv,delta_u_conv} into \cref{eq:gain_norm}, we obtain the same form, but with $\delta u^{\rm conv}$ in the denominator: $\Delta \tilde{\varepsilon} \approx \sqrt{2}\pi / \sqrt{\delta u^{\rm conv}}$. Since $\delta u^{\rm conv} \gtrsim \delta u^{\rm str}$, the energy gain is reduced.

Our finding then is that the channel wall tilt is detrimental. If the tilt is sufficiently strong to satisfy the condition given by \cref{slope_cond}, the reflection contribution becomes important. It causes the longitudinal electric field to increase. However, this also causes the slip determined by $\delta u$ to increase. The increase of $\delta u$ outweighs any benefit from increasing $E_x$.

It is now worth reexamining our simulation results with the help of these estimates by first focusing on the field structure. By comparing \cref{fig:Ex_phase}(b) and \cref{fig:Ex_phase}(d), we find that $\delta u$, which characterizes the tilt of the wavefronts in these plots, is bigger for $E_x$ than for $E_y$ by about a factor of two. This is a clear indicator that we have multiple modes or waves in the channel, because otherwise $\delta u$ would be the same for $E_x$ and $E_y$. A similar discrepancy was reported in Ref.~[\onlinecite{Willingale.NJP.2018.Longitudinal}] where it was shown that it can be produced by overlapping a strong main wave propagating along the axis with two weaker waves propagating at an angle. In that picture, the main wave dominates $E_y$, while $E_x$ is generated by the weaker, oblique waves. In our case, the main wave would be the fundamental mode of the straight channel, so its phase velocity corresponds to $\delta u^{\rm str}$. On the other hand, the oblique waves are produced by the reflection off the tilted channel wall, so their phase velocity corresponds to $\delta u^{\rm conv}$. Since $E_y$ is dominated by the main wave while $E_x$ is dominated by the oblique waves, $\delta u$ inferred from $E_y$ is close to $\delta u^{\rm str}$, while $\delta u$ inferred from $E_x$ is close to $\delta u^{\rm conv}$.

We now use the simulation results to estimate the two values of $\delta u$. According to \cref{fig:current_and_Ex}, the channel radius at $x/\lambda_0 \approx 50$ is roughly $r_{\rm ch} \approx 5\lambda_0$. The phase velocity of $E_y$ in \cref{fig:Ex_phase} gives $\delta u^{\rm str} \approx 0.004$. Using \cref{delta_u_str}, we find that $\alpha \approx 2.8$, an order-unity value, as expected for a fundamental guided mode. With this value of $\alpha$, the condition given by \cref{slope_cond} for the reflection contribution to dominate becomes $|\Delta r_{\rm ch}/\Delta x| \gtrsim 0.045$. The wall of the channel in \cref{fig:plasma_current} over $x/\lambda_0 \approx 50$--$65$ has a tilt of $|\Delta r_{\rm ch}/\Delta x| \approx 0.06$, so this condition is indeed satisfied, and the effect of the tilt is not negligible. Using this tilt in \cref{delta_u_conv} and \cref{E_x_conv}, we obtain $\delta u^{\rm conv} \approx 0.007$ and $E_x^{\rm conv}/E_0 \approx 0.12$. These values are in reasonably good agreement with the phase velocity and amplitude of $E_x$ measured directly in \cref{fig:Ex_phase}(d), confirming the consistency between our estimates and the simulation results.

Finally, we reexamine \cref{fig:energy_vs_time} and \cref{fig:energy_accounting} with the help of the derived estimates. \Cref{fig:energy_vs_time} gives the time evolution of the electron energy-weighted distribution. The energy of the most energetic population is consistent with our estimate for energy gain from the longitudinal field. This population emerges by $t \approx 100 \tau$. This time scale is consistent with the time required to reach the peak energy while surfing with the longitudinal field, which we denoted as $\Delta t$. Using the value of $\delta u$ measured for $E_x$ from \cref{fig:Ex_phase}(d), $\delta u \approx 0.008$, we obtain $\Delta t/\tau \approx 1/(2 \times 0.008) \approx 60$.

\Cref{fig:scatter} provides a helpful illustration of the process. It shows a random 1-in-200 sample of energetic electrons with $\varepsilon>a_0mc^2$ in the polarization and orthogonal slabs at three different times. In the polarization slab, we see energetic electrons inside the channel near the wall, consistent with acceleration as they move forward along the channel wall. The orthogonal slab does not show a similar energetic population. We also see that energetic electrons return into the surrounding plasma. The number of energetic electrons grows, which is consistent with the evolution of the electron spectrum in \cref{fig:energy_vs_time}(a).

The continuous and strong wave depletion in \cref{fig:energy_accounting} is therefore consistent with the discussed mechanism. Newly injected particles continue to deplete the wave, while the retention of energy by particles that have already returned to the surrounding plasma is what supports the growing energetic population. It is worth pointing out that the injection of new particles is critical for preventing saturation. This is why the single-particle acceleration time $\Delta t$ does not appear as a saturation time in \cref{fig:energy_accounting}.

\section{Summary and discussion}
\label{sec:summary}

Our simulation demonstrates how a relativistically strong electromagnetic wave of finite transverse width interacting with a cold pair plasma creates a fully evacuated channel, without penetrating the plasma. For the parameters considered here, the plasma is opaque according to the relativistic plane-wave transparency criterion of Ref.~[\onlinecite{Tangtartharakul.PRL.2026.Reflection}], with $\omega_p/\omega_0$ well above the threshold $\omega_p/\omega_0 \approx 1/\sqrt{32}a_0$, making channel formation unavoidable. It is plausible that similar channels also form in different regimes, in particular through the expected filamentation instability.

The main surprising feature of the wave-plasma interaction observed in the simulation is that the wave loses energy much faster than one would expect based on the cost of plasma clearing. Remarkably, the interaction is accompanied by strong particle acceleration. \AVA{Our analysis shows that the particles are accelerated by the longitudinal electric field associated with the transverse structure of the wave and the geometry of the self-generated channel. \AVA{Because of their opposite charges, electrons and positrons surf in opposite phases of this field: electrons are accelerated in regions with $E_x<0$, while positrons are accelerated in regions with $E_x>0$. This species separation and phase alignment allow both species to gain energy while moving forward along the channel walls for a time much longer than the wave period.} In our simulation, the resulting energetic particles are found to carry most of the energy lost by the wave.}

The evacuated channel formed by the wave has a direct analogy to hollow-core targets used in experiments with ultra-high-intensity lasers. These targets similarly guide the laser and suppress diffraction while maintaining a strong longitudinal electric field over distances much longer than the vacuum Rayleigh range. Electron acceleration in such targets has been studied using numerical simulations and analytical theory~\cite{Gong.PPCP.2019.Hollow,Wang.PRE.2021.Dimensionality,Xiao.PRE.2016.Longitudinal}. The acceleration mechanism detailed here is closely related to that at work in hollow-core targets, but in our case the channel forms self-consistently through the interaction of a wave of finite transverse width with the pair plasma. \AVA{There is, however, a difference in the generated magnetic-field structure. In the hollow-core target, the current is carried only by electrons~\cite{Gong.PPCP.2019.Hollow}, whereas in our case it is carried by both electrons and positrons. The resulting alternating current in the pair plasma generates an alternating rather than a quasi-static azimuthal magnetic field.}

The interaction of strong electromagnetic waves with pair plasma may have important applications to FRBs. In particular, damping of strong radio waves during their propagation through the ambient $e^\pm$ plasma was found to provide significant constraints on FRB models~\cite{Beloborodov.APJ.2024.Damping,Beloborodov.APJ.2026.Compression}. Existing calculations of FRB damping are limited to plane-wave geometry while it is known that FRBs are prone to filamentation instability~\cite{Sobacchi.APJL.2023.Filamentation}. Since filamentation clears out low-density channels in the plasma, one might expect that it helps the radio wave avoid damping and escape. Our results raise the possibility that filamentation can instead increase the damping rate by enabling fast particle heating near the walls of the channel.

The wave-plasma interaction regime simulated in the present paper differs from FRB scenarios in which the radio wave propagates through the surrounding pair plasma. In our simulations, $\omega_p/\omega_0$ is well above the transparency ceiling $\omega_p/\omega_0 \approx 1/(\sqrt{32}a_0)$ given by Eq.~(\ref{eq:1D_threshold}) according to the relativistic plane-wave criterion of Ref.~[\onlinecite{Tangtartharakul.PRL.2026.Reflection}]. Furthermore, our simulation setup did not include strong background magnetic fields $B_{\rm bg}$ that exist around neutron stars and play a key role for FRBs in a few ways. (i) The large parameter $\sigma=B_{\rm bg}^2/4\pi n_0 m_e c^2\gg 1$ makes the magnetized plasma effectively heavier by the factor of $\sigma$, which delays its bulk acceleration by the wave~\cite{Beloborodov.APJ.2026.Compression}. (ii) A magnetic field $B_{\rm bg}>E_0$ can allow orthogonally polarized waves $\boldsymbol{E}\perp \boldsymbol{B}_{\rm bg}$ to propagate through the dense plasma~\cite{Canuto.PRD.1971.Scattering}, although other processes can still strongly damp the wave~\cite{Golbraikh.APJ.2023.Escape,Solanki.ArXiv.2026.Damping}. (iii) A relatively small $B_{\rm bg}<E_0$ can enable fast stochastic heating of the plasma by the wave, leading to enormous radiative losses and fast damping of the wave~\cite{Beloborodov.PRL.2022.Scattering}.

Wave filamentation is still expected in these strongly magnetized plasmas, and the effect of particle acceleration reported in the present paper may not be specific to the unmagnetized case. We find that particle acceleration accompanies the sideways expulsion of the plasma by the ponderomotive force. A similar mechanism may operate in a strongly magnetized plasma, although the expulsion would be expected to occur preferentially along $\boldsymbol{B}_{\rm bg}$, producing a layered rather than conical plasma structure. We leave the investigation of these effects in the presence of a strong $B_{\rm bg}$ to future work.

\section*{Acknowledgments}

The research presented here was supported by the National Science Foundation (PHY-2512067, PHY-2206777) and by the Air Force Office of Scientific Research (FA9550-24-1-0053). A.M.B. acknowledges support by NASA grant 80NSSC24K1229, NSF grant AST-2408199, and Simons Foundation grant 446228. This work was facilitated by Multimessenger Plasma Physics Center (MPPC) grant PHY-2206609. This work used Stampede3 at TACC through allocation TG-PHY250187 and TG-PHY250225 from the Advanced Cyber Infrastructure Coordination Ecosystem: Services \& Support (ACCESS) program, which is supported by National Science Foundation grants \#2138259, \#2138286, \#2138307, \#2137603, and \#2138296. Simulations were performed using EPOCH, which was developed as part of the UK EPSRC-funded projects no. EP/G054950/1, EP/G056803/1, EP/G055165/1 and EP/M022463/1.

\section*{References}
\bibliographystyle{apsrev4-1}
\bibliography{Collections}

\appendix

\section{1D relativistic reflection front sustained by a linearly polarized wave}
\label{sec:app:rel-hole-boring-lp}

We follow a standard momentum-balance analysis~\cite{Robinson.PPCF.2009.Boring} to derive a solution for a relativistic reflection front sustained by a plane linearly polarized electromagnetic wave irradiating a cold pair plasma. A similar derivation for a circularly polarized wave was given in Ref.~[\onlinecite{Tangtartharakul.PRL.2026.Reflection}]. Here we repeat the analysis for a linearly polarized wave in order to obtain the correct numerical factors for the expressions used in the main text.

The reflection front is maintained by an incident linearly polarized wave of constant normalized amplitude $a_0$. This boundary sweeps up all of the pair plasma while moving with average velocity $v_b$. It is therefore convenient to examine the front in a frame moving with velocity $v_b$, where the front is stationary on average. We denote quantities in this co-moving frame using a prime. For compactness, we also introduce the normalized velocity of the front $\beta_b \equiv v_b/c$.

In the co-moving frame, the problem reduces to a momentum-balance condition averaged over one wave oscillation. The electromagnetic wave transfers momentum to the plasma by reflecting from the front. In our model, this cycle-averaged electromagnetic momentum flux is balanced by the momentum flux of the pair-plasma particles swept up by the front. In the co-moving frame, this particle flux results from the incoming plasma flow being reflected by the front. To simplify the problem, we treat the wave and particle reflection as elastic.

The first step is to obtain the expression for the electromagnetic momentum flux. The incident wave is linearly polarized, so the momentum flux oscillates and must be averaged over one wave period. For elastic reflection, the cycle-averaged flux to the left of the front is
\begin{equation}
-\frac{\langle E'^2\rangle}{2\pi} =
-\frac{E_0'^2}{4\pi},
\end{equation}
where
\begin{equation}
    E_0'=(a_0 m_e c/|e|)\omega_0'
\end{equation}
is the peak electric field and
\begin{equation}
\omega_0'=\omega_0 \sqrt{\frac{1-\beta_b}{1+\beta_b}}
\end{equation}
is the Doppler-shifted frequency.

The next step is to find the momentum flux of the plasma. In the co-moving frame, the plasma that has not interacted with the wave flows toward the front with speed $\beta_b c$. The density of each species in this flow is $n'=\gamma_b n_0$, where
\begin{equation}
\gamma_b=\frac{1}{\sqrt{1-\beta_b^2}} .
\end{equation}
Since the particles are elastically reflected, each species contributes twice its incoming momentum flux. Accounting for both species, the total plasma momentum flux to the right of the front is $-4 n'\gamma_b m_e c^2\beta_b^2$.

The momentum-balance condition is obtained by equating the electromagnetic flux to the plasma flux:
\begin{equation}
a_0^2 n_{\rm cr}\frac{1-\beta_b}{1+\beta_b}
= 4 n_0\gamma_b^2\beta_b^2 .
\label{eq:pair-front-balance-lp}
\end{equation}
This equation determines the normalized velocity of the reflection front, $\beta_b$. Introducing the parameter
\begin{equation}
\psi_{\rm LP}
\equiv \frac{4 n_0}{a_0^2 n_{cr}},
\end{equation}
the momentum-balance equation reduces to
\begin{equation}
(1-\beta_b)^2=\psi_{\rm LP}\beta_b^2 .
\end{equation}

The physically relevant solution corresponds to a front moving into the plasma. We therefore have
\begin{equation}
\beta_b = \frac{1}{1+\sqrt{\psi_{\rm LP}}}.
\label{eq:beta-b-lp}
\end{equation}
This expression has the same form as the circularly polarized result obtained in Ref.~[\onlinecite{Tangtartharakul.PRL.2026.Reflection}]. The difference is that the parameter denoted here as $\psi_{\rm LP}$ is twice as large as the corresponding circular-polarization parameter,
\begin{equation}
\psi_{\rm LP}=2\psi_{\rm CP},
\end{equation}
where $\psi_{\rm CP}=2 n_0/a_0^2 n_{cr}$. This factor of two reflects the smaller cycle-averaged radiation pressure of a linearly polarized wave with the same peak amplitude $a_0$.

We can now use the obtained result to find the particle density at the moving front in the laboratory frame of reference. In the co-moving frame, the total density is $2n'$ for each species, representing the sum of the incoming and reflected flows. We need to transform the density of each flow back to the laboratory frame. In the laboratory frame, the incoming flow is simply the immobile plasma ahead of the front, so its density for each species is $n_0$. The reflected electrons and positrons have density $n'=\gamma_b n_0$ and speed $\beta_b c$ in the co-moving frame. Transforming this density back to the laboratory frame gives
\begin{equation}
n_{\rm ref} = \gamma_b n'(1+\beta_b^2)
=
(2\gamma_b^2-1)n_0 .
\end{equation}
This is the density of the swept-up particles of each species associated with the reflected flow in the co-moving frame. The total density for each species at the moving front is then
\begin{equation}
n_{\rm plug} = n_0+n_{\rm ref} = 2\gamma_b^2 n_0 .
\label{eq:plug-density-lp}
\end{equation}
We denote this density by $n_{\rm plug}$ because it corresponds to the dense plug produced at the front.

\begin{table}
\centering
\renewcommand{\arraystretch}{1.18}
\begin{tabular}{|p{3.7cm}|P{4.6cm}|}
\hline
\multicolumn{2}{|c|}{Incident wave parameters} \\
\hline
Normalized amplitude & $a_0 = 50$ \\
\hline
Propagation direction & $+x$ \\
\hline
Polarization & linear, $\bm{E} \parallel \bm{e}_y$ \\
\hline
Focal position & $x_\mathrm{focus}=0$ \\
\hline
Waist size & $w_0 = 5.0\lambda_0$ \\
\hline
Pulse duration & $\tau_{\mathrm{FWHM}} = 37.0\lambda_0/c$ \\
\hline
Peak time & $t_\mathrm{peak} = 92.5\lambda_0/c$ \\
\hline
\hline
\multicolumn{2}{|c|}{Pair plasma parameters} \\
\hline
Composition & electrons and positrons \\
\hline
Density of each species & $n_0 = 10 n_{cr}$ for $x \ge 0$ \\
\hline
Density ramp & linear, from $n_0/20$ to $n_0$ over $-6.25 < x/\lambda_0 < 0$ \\
\hline
Target dimensions & $617\lambda_0 \times 50\lambda_0 \times 50\lambda_0$ \\
\hline
Grid resolution & $16.2$ cells/$\lambda_0$ in each direction \\
\hline
Macroparticles per cell & 10 per species \\
\hline
\end{tabular}
\caption{Normalized input parameters of the 3D PIC simulation.}
\label{table:pic-norm}
\end{table}

\section{Parameters used in PIC simulations}
\label{sec:setup}

The normalized parameters used in the 3D PIC simulation are listed in \cref{table:pic-norm}. The simulation was performed using the fully relativistic PIC code EPOCH~\cite{Arber.PPCF.2015.EPOCH}. While the physics is governed by normalized quantities, EPOCH requires dimensional quantities as input. To make it easier to reproduce our results, we provide the dimensional parameters used to perform the simulation and additional numerical settings in \cref{table:pic-dim}.

The incident wave is a linearly polarized Gaussian beam propagating in the positive direction along the $x$-axis. The beam parameters are given in \cref{table:pic-norm,table:pic-dim}. In the focal plane located at $x=x_\mathrm{focus}$ and in the absence of plasma, the intensity is
\begin{equation}
I = I_0\exp\left[-\frac{2(y^2+z^2)}{w_0^2}-4\ln 2\frac{(t-t_\mathrm{peak})^2}{\tau_\mathrm{FWHM}^2}\right].
\end{equation}
The beam is launched from a boundary of the simulation domain located in vacuum. The acceleration mechanism discussed in the main text is sensitive to the phase velocity of the wave. To reduce numerical dispersion, we use a fourth-order finite-difference time-domain field solver rather than a conventional second-order scheme~\cite{Yee.IEEE.1966.FDTD,Young.IEEE.1997.4FDTD}.

The wave irradiates an initially cold electron-positron plasma slab with equal electron and positron densities, $n_e=n_p=n_0$. The uniform part of the target begins at $x=0$, where the density reaches $n_0=10n_{cr}$. The focal plane is located at the front edge of the uniform plasma. A linear density ramp is placed at the front of the target, with the density increasing from $n_0/20$ to $n_0$ over $-6.25 < x/\lambda_0 < 0$.

\begin{table}
\centering
\renewcommand{\arraystretch}{1.18}
\begin{tabular}{|p{3.9cm}|P{4.8cm}|}
\hline
\multicolumn{2}{|c|}{Incident laser parameters} \\
\hline
Normalized amplitude & $a_0 = 50$ \\
\hline
Wavelength & $\lambda_0 = 0.81~\micron$ \\
\hline
Period & $\tau = 2.71~{\rm fs}$ \\
\hline
Propagation direction & $+x$ \\
\hline
Polarization & linear, $\bm{E} \parallel \bm{e}_y$ \\
\hline
Electric-field amplitude & $E_0 = 2.0\times 10^{14}~{\rm V/m}$ \\
\hline
Magnetic-field amplitude & $B_0 = 7.0\times 10^5~{\rm T}$ \\
\hline
Peak intensity & $I_0 = 5.2\times 10^{21}~{\rm W/cm^2}$ \\
\hline
Cycle-averaged peak power & $0.84~{\rm PW}$ \\
\hline
Total laser energy & $156~{\rm J}$ \\
\hline
Focal position & $x_\mathrm{focus}=0$ \\
\hline
Waist size & $w_0 = 4.0~\micron$ \\
\hline
Pulse duration & $\tau_{\mathrm{FWHM}} = 100~{\rm fs}$ \\
\hline
Peak time & $t_\mathrm{peak}=250~{\rm fs}$ \\
\hline
\hline
\multicolumn{2}{|c|}{Pair plasma parameters} \\
\hline
Composition & electrons and positrons \\
\hline
Critical density & $n_{cr}=1.7\times 10^{21}~{\rm cm^{-3}}$ \\
\hline
Density of each species & $n_0 = 10 n_{cr}$ for $x \ge 0$ \\
\hline
Density ramp & linear, from $n_0/20$ to $n_0$ over $-5~\micron <x<0$ \\
\hline
Target dimensions & $75~\micron \times 40~\micron \times 40~\micron$ \\
\hline
\hline
\multicolumn{2}{|c|}{Numerical parameters} \\
\hline
Field solver & fourth-order FDTD \\
\hline
CFL constant & 0.98 \\
\hline
Grid resolution & 20 cells/$\micron$ in each direction \\
\hline
Macroparticles per cell & 10 per species \\
\hline
\end{tabular}
\caption{Dimensional parameters and numerical settings of the 3D PIC simulation.}
\label{table:pic-dim}
\end{table}

%**************************************

\section{Comparison with circular polarization}
\label{app:circular-polarization}

To show that the main physical picture described in the main text is not sensitive to the polarization of the wave, we performed an additional simulation using a circularly polarized (CP) pulse. This Appendix presents the results of that simulation and compares them with those of the linearly polarized (LP) simulation described in the main text.

We use the same simulation setup and only adjust the pulse. The CP pulse has the same total energy as the LP pulse,
\begin{equation}
    \mathcal{E}_{\mathrm{pulse,CP}}
    = \mathcal{E}_{\mathrm{pulse,LP}}
    = 156~\mathrm{J}.
\end{equation}
The normalized amplitude was therefore reduced to $a_{0,\mathrm{CP}}=a_{0,\mathrm{LP}}/\sqrt{2} \approx 35.4$ to keep the total pulse energy unchanged. To limit the computational cost, we used three electron macroparticles and three positron macroparticles per cell.

\begin{figure}
    \centering
    \includegraphics[width=0.94\linewidth]{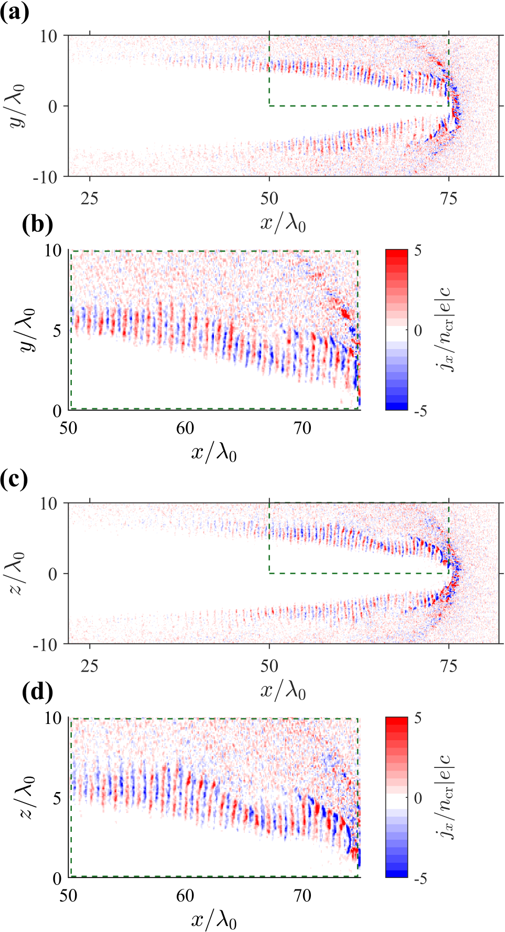}
    \caption{Longitudinal plasma current density $j_x$ in the CP simulation at $t/\tau=148$. Panels (a) and (b) show the $(x,y)$-plane at $z=0$, with panel (b) showing a zoomed-in view of the dashed region in panel (a). Panels (c) and (d) show the $(x,z)$-plane at $y=0$, with panel (d) showing a zoomed-in view of the dashed region in panel (c).}
    \label{fig:app:cp-current}
\end{figure}

\begin{figure}
    \centering
    \includegraphics[width=0.8\linewidth]{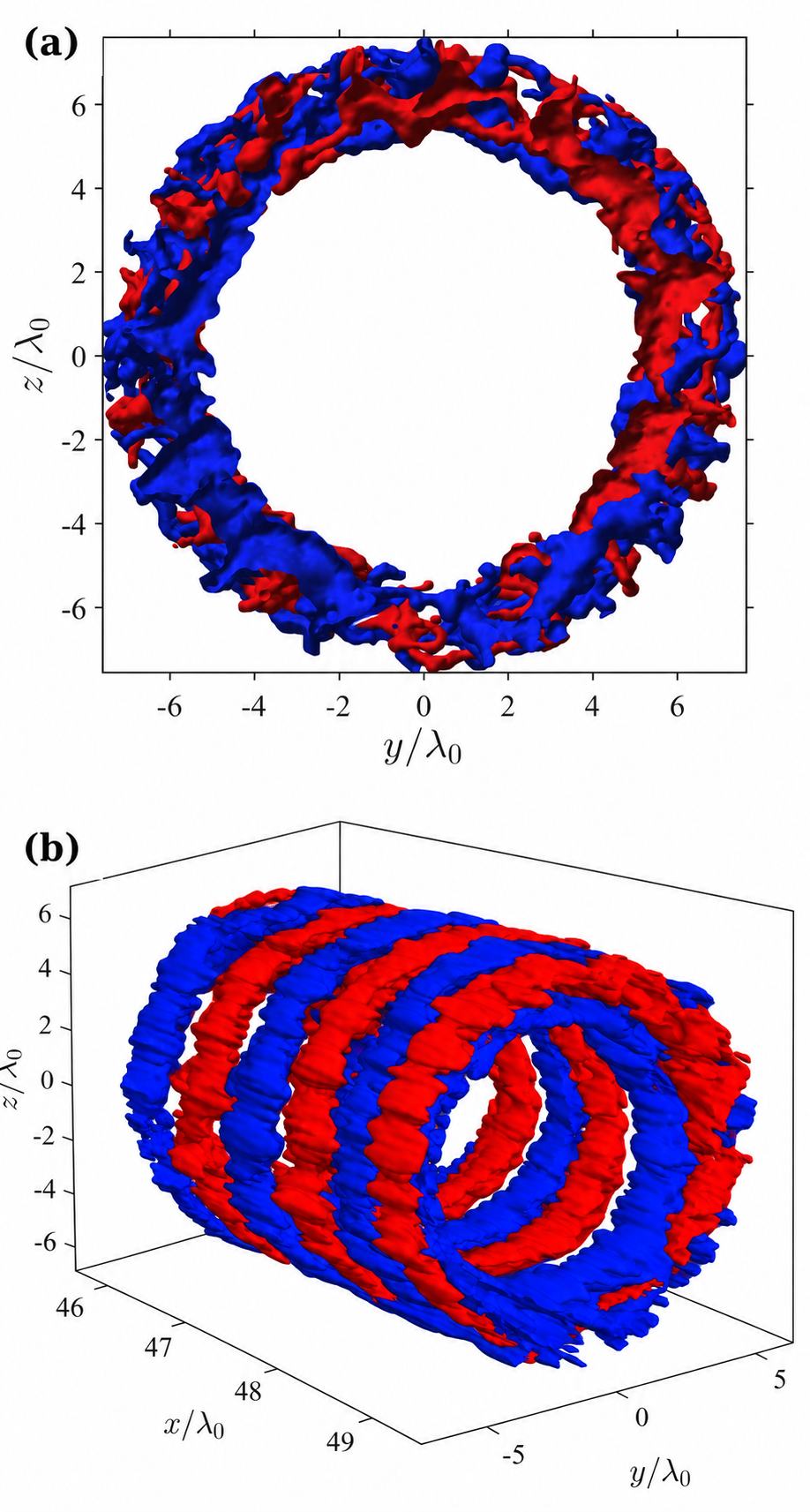}
    \caption{Longitudinal current density at $t/\tau=148$, shown using the two isosurfaces $j_x/(n_{cr}|e|c) = \pm 0.5$. In both panels, only the region $46 \le x/\lambda_0 \le 49$ is shown to improve visibility. Panels (a) and (b) show head-on and oblique views, respectively.}
    \label{fig:app:cp-current-3D}
\end{figure}

\Cref{fig:app:cp-current} shows snapshots of the longitudinal current $j_x$ in two orthogonal slices. The key difference from the LP case shown in \cref{fig:plasma_current} is that there is now also strong longitudinal current in the $(x,z)$-plane. The underlying reason is that there is no preferred transverse direction for circular polarization and the particle injection is distributed around the entire periphery of the channel rather than being localized in the polarization plane. \Cref{fig:app:cp-current-3D} shows this structure more directly using three-dimensional isosurfaces of $j_x$.

\begin{figure}
    \centering    \includegraphics[width=0.9\linewidth]{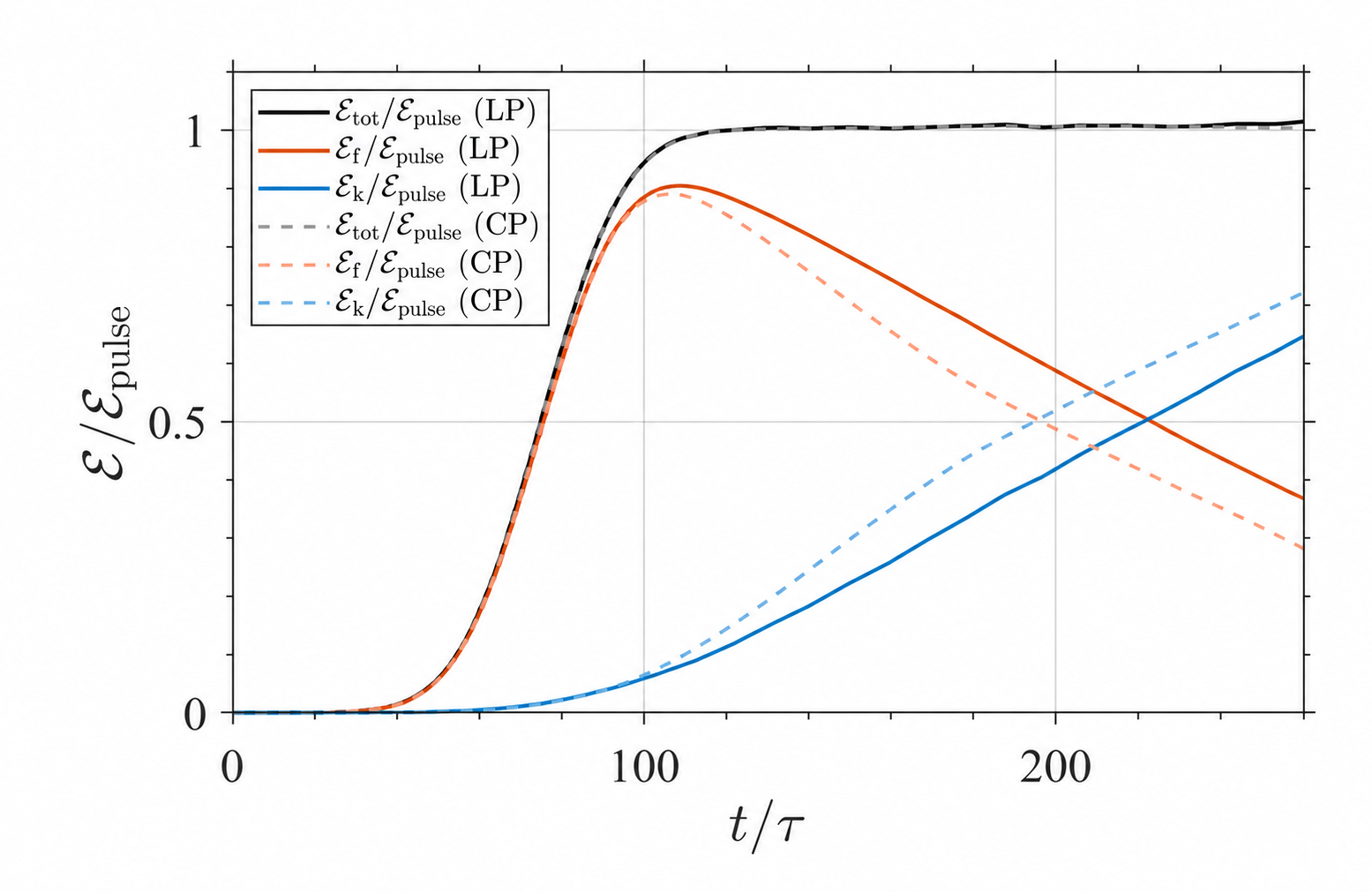}
    \caption{Energy accounting for the LP (solid) and CP (dashed) simulations. The curves show the temporal evolution of the field energy $\mathcal{E}_{\mathrm{f}}$ (red), particle kinetic energy $\mathcal{E}_{\mathrm{k}}$ (blue), and total energy $\mathcal{E}_{\mathrm{tot}}=\mathcal{E}_{\mathrm{f}}+\mathcal{E}_{\mathrm{k}}$ (black), normalized to the full energy in the incident wave $\mathcal{E}_{\mathrm{pulse}}$.}
    \label{fig:app:cp-energy}
\end{figure}

\Cref{fig:app:cp-energy} compares the energy accounting for the CP and LP simulations. The blue curves show the energy transferred to the particles, while the red curves show the remaining field energy. A direct comparison shows that the field-to-particle energy transfer and wave depletion are similar in the two cases. Together with the current distributions in \cref{fig:app:cp-current}, this shows that the main physical picture described in the main text remains qualitatively unchanged for circular polarization.

%The CP energy accounting in \cref{fig:app:cp-energy}, shown together with the LP curves, follows the same overall pattern as the LP result in \cref{fig:energy_accounting}. In both simulations, the black total-energy curve remains close to unity, the red field-energy curve decreases after the pulse enters the plasma, and the blue particle-energy curve increases. Over their common time interval, the CP traces therefore exhibit the same qualitative field-to-particle energy transfer and wave depletion as the LP traces. Together with the current distributions in \cref{fig:app:cp-current}, these results show qualitatively similar injection, current formation, and energy-depletion behavior for the CP and LP simulations.

%**************************************

\section{Generation of the wave-field overlay}
\label{app:field-overlay}

\begin{figure}[tb]
    \centering
    \includegraphics[width=1.0\linewidth]{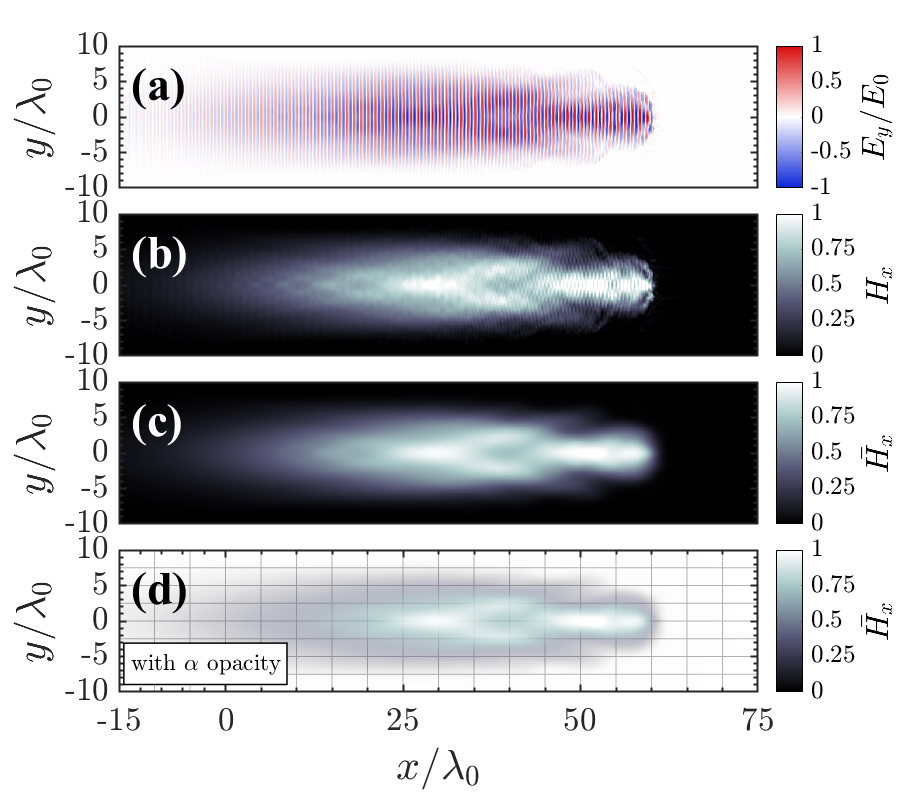}
    \caption{Generation of the wave-field overlay from the $(x,y)$ slice of $E_y$. (a) Raw normalized field $E_y/E_0$. (b) Envelope field $H_x=\left|\mathcal{H}_x\{E_y/E_0\}\right|$ obtained from the analytic signal. (c) Normalized smoothed envelope $\bar{H}_x$. (d) Transparent rendering of the same normalized envelope, with opacity $\alpha=\bar{H}_x$. The background grid in (d) is included only to illustrate the opacity of the overlay.}
    \label{fig:overlay}
\end{figure}

The wave-field overlay used in Fig.~\ref{fig:density} was obtained from the transverse electric field $E_y$ in the $(x,y)$ plane. The purpose of the overlay is to show the slowly varying envelope while suppressing the rapid sign changes of the raw oscillating field. The procedure is summarized in Fig.~\ref{fig:overlay}.

We begin with the normalized field
\begin{equation}
    f(x,y) = E_y(x,y)/E_0,
\end{equation}
shown in Fig.~\ref{fig:overlay}(a), where $E_0$ is the peak amplitude in the absence of the plasma. Since the field oscillates primarily along the propagation direction, the envelope is computed independently along each line of constant $y$. Let $\mathscr{H}_x[f]$ denote the ordinary one-dimensional Hilbert transform of $f$ along $x$. We use it to define the analytic signal
\begin{equation}
    \mathcal{H}_x\{f\}(x,y) \equiv f(x,y) + i\mathscr{H}_x[f](x,y),
\end{equation}
and the corresponding envelope field
\begin{equation}
    H_x(x,y) \equiv \left|\mathcal{H}_x\{f\}(x,y)\right|.
    \label{eq:hx_def}
\end{equation}
This operation removes the rapidly changing sign of the field and returns a positive local amplitude, as shown in Fig.~\ref{fig:overlay}(b).

To remove residual short-scale structure and produce a smooth overlay, $H_x$ is smoothed with a Gaussian filter and normalized to its maximum value in the plotted $(x,y)$ slice:
\begin{equation}
    \bar{H}_x(x,y) \equiv \frac{\mathcal{G}_{\lambda_0}\!\left[H_x\right](x,y)}{\max_{x,y}\mathcal{G}_{\lambda_0}\!\left[H_x\right](x,y)}.
    \label{eq:gaussian_smoothing}
\end{equation}
Here $\mathcal{G}_{\lambda_0}$ denotes Gaussian smoothing with widths $\sigma_x=\lambda_0$ and $\sigma_y=0.5\lambda_0$. Equivalently, this smoothing can be viewed as a convolution of $H_x$ with an anisotropic Gaussian kernel. The smoothing is useful because the envelope field can still contain residual short-scale modulations, especially near the pulse front and in regions where diffraction, interference, or numerical noise introduce fine structure. The normalized smoothed envelope $\bar{H}_x$ is shown in Fig.~\ref{fig:overlay}(c).

The same normalized scalar field is used both as the grayscale overlay value and as the opacity,
\begin{equation}
    \alpha(x,y) = \bar{H}_x(x,y).
    \label{eq:overlay_alpha}
\end{equation}
Thus, regions with small envelope amplitude are nearly transparent, while regions with large envelope amplitude are nearly opaque. This rendering step is illustrated in Fig.~\ref{fig:overlay}(d), where a background grid is shown only to demonstrate the opacity mapping.

\end{document}